\documentclass[11pt]{article}
\usepackage{longtable}
\usepackage{lscape}
\usepackage{graphicx}
\usepackage[super,sort&compress,comma]{natbib} 
\usepackage{amsmath}

\usepackage{setspace}

\title{Atomistic details of oxide surfaces and surface oxidation: the example of copper and its oxides}
\author{Chiara Gattinoni and Angelos Michaelides\\Thomas Young Centre, London Centre for Nanotechnology and Department of Chemistry\\ University College London\\17-19 Gordon Street, London, WC1H 0AH, UK}
\date{}

\begin{document} 
\maketitle


\begin{abstract}
The oxidation and corrosion of metals are fundamental problems in materials science and technology that have been studied using a large variety of experimental and computational techniques. 
Here we review some of the recent studies that have led to significant advances in our atomic-level understanding of copper oxide, one of the most studied and better understood metal oxides.
We show that a good atomistic understanding of the physical characteristics of cuprous (Cu$_2$O) and cupric (CuO) oxide and of some key processes of their formation has been obtained.
Indeed, the growth of the oxide has been proven to be epitaxial with the surface and to proceed, in most cases, through the formation of oxide nano-islands which, with continuous oxygen exposure, grow and eventually coalesce.
We also show how electronic structure calculations have become increasingly useful in helping to characterise the structures and energetics of various Cu oxide surfaces.
However a number of challenges remain.
For example, it is not clear under which conditions the oxidation of copper in air at room temperature (known as native oxidation) leads to the formation of a cuprous oxide film only, or also of a cupric overlayer.
Moreover, the atomistic details of the nucleation of the oxide islands are still unknown.
We close our review with a perspective on future work and discuss how recent advances in experimental techniques, bringing greater temporal and spatial resolution, along with improvements in the accuracy, realism and timescales achievable with computational approaches make it possible for these questions to be answered in the near future.

\end{abstract}

\section{Introduction}
\label{intro}

Copper is a material which has accompanied human pre-history and history, and it is still highly relevant today.
Cold working of copper has been performed for at least $10,000$ years and smelting of copper ore for around $7,000$ years~\cite{miodownik}.
Its use as construction materials, \emph{e.g.} in piping, can be dated back to ancient Egypt, and its importance in this field has not diminished nowadays.
In the modern world it has acquired further uses, for example in electrical systems and electronic devices.

Within this long history, the properties of copper have been extensively studied and exploited, however much is still unknown about this important metal.
In particular, the oxidation and corrosion of copper, which impacts its performance in industrial and technological applications, is still not completely understood.

Copper is found to readily oxidise at room temperature~\cite{barr1,iijima,platzman_2008}, and the presence of an oxide layer, however thin, can compromise its uses in technology.
As an example, copper could be an environmentally friendly and low-cost substitute for the (currently used) tin-lead or (promising) gold- and silver-based solder alloys in electronic packaging, if there was a way to prevent its oxidation~\cite{song_metals_2012}.
Moreover, copper canisters are used for nuclear waste disposal, and understanding the oxidation and corrosion of copper in anaerobic conditions is thus really important~\cite{nuclear}.
On the other hand, the existence of stable copper oxides at room temperature, with a $\sim 2.0$ eV band gap, makes them interesting for catalytic~\cite{reitz_solomon}, gas sensing~\cite{nakamura_jes_1990}, optoelectronic and solar technologies~\cite{Rai1988265, meyer}.
Thus, there currently is a two-fold interest in understanding copper oxides: from the one hand, to mitigate against technological failure, on the other hand, to exploit their potential industrial applications.

Copper is considered a model system to understand the formation of metal oxides in general.
The atomistic details of the oxidation process tend to be system-specific, with some metals showing uniform oxide growth (\emph{e.g.} Ref.~\cite{alumina}), other complex temperature-dependent phenomena such as surface roughening (\emph{e.g.} Ref.~\cite{silica}) and island formation (\emph{e.g.} Ref.~\cite{chromium, yang_apl_2002}).
However, the copper oxidation process is one of the most studied with a large number of experimental and computational methods, and one of the better understood.
Therefore, a detailed understanding of copper oxidation, of the techniques used to study it and of the challenges which are still open is invaluable when considering the oxidation process on any other system.

In this review, we discuss the status of knowledge of copper oxidation from the atomistic point of view, which we believe is of key importance if we want to learn how to prevent or manipulate copper oxidation.
We cannot hope to provide a complete review of all the work done on this subject since the beginning of the last century~\cite{lawless, freund, diebold}.
We are therefore only going to focus on recent surface science, spectroscopy and atomistic computational work which has been performed to understand the properties of copper oxides and their formation, and on the open challenges that can be addressed using these techniques.

First, a brief overview on the experimental and computational techniques which have been used for the study of oxide structures and oxidation kinetics is given (Sec.~\ref{techniques}), in order to clarify some of the terminology used throughout the review.
The structural and electronic characteristics of the bulk oxides and their surfaces are then presented (Sec.~\ref{oxide_structures}).
We subsequently look at the interaction of clean copper surfaces with oxygen and the initial stages of controlled oxidation (Sec.~\ref{o_ads}) as well as long-term oxidation (Sec.~\ref{controlled}).
Finally, in Sec.~\ref{conclusions}, conclusions and perspectives are given.

We hope that it will be clear from the following that tremendous progress has been made in understanding the atomistic details of copper oxides and their formation under different conditions.
However, equally important gaps in our understanding remain, especially in terms of the formation kinetics and the structure of the resulting oxide surfaces.

\section{Experimental and computational techniques}
\label{techniques}

An enormous number of experimental~\cite{somorjai, gross_book, wandelt, bracco2013surface} and computational techniques are available to investigate the physical and chemical characteristics of solids, surfaces and surface kinetic processes.
Many of them have been used over the years to \emph{e.g.}, understand oxide structures, characterize oxide surfaces, understand the oxidation kinetics and investigate bulk properties of the oxides.
For clarity, in this section we provide a brief introduction to the most relevant techniques which have been used on copper oxide and which are going to be mentioned in the following sections, with an emphasis on strengths and weaknesses of each approach.

In early studies of oxidation, thermogravimetric analysis (TGA), where changes in physical and chemical properties of materials are measured as a function of time, has been widely used to study the onset of oxidation by recording the mass gain of a sample under oxygen exposure.
Whilst useful in providing a broad overview of the extent of oxidation, this technique is however unable to provide atom-resolved information.

Imaging techniques such as electron microscopy (EM), transmission electron microscopy (TEM) and their derivatives (\emph{e.g.}, high-resolution transmission electron microscopy (HRTEM) or field emission scanning electron microscopy (FESEM)) and surface-specific techniques such as scanning tunnelling microscopy (STM) can be applied to surfaces and provide atomistic level structural information.
As we will see, they have been amply used to image adsorption of oxygen on the copper surface, surface reconstructions and initial oxide formation.
The atomic composition and oxidation states of the atoms in a material can be obtained using spectroscopic techniques such as X-ray diffraction (XRD), Auger electron spectroscopy (AES), electron energy loss spectroscopy (EELS), low-energy electron diffraction (LEED), reflection high-energy electron diffraction (RHEED), X-ray photoelectron spectroscopy (XPS or ESCA) and X-ray absorption spectroscopy (XAS).
The space- and time-resolution of these techniques has greatly improved since they were first used in this field (in the 1980s), as well as their range of applicability: for example, experiments at relatively high pressures can be performed nowadays~\cite{salmeron_xas}.

Growth of the oxide and its atomic composition has been extensively studied by means of ellipsometry.
This technique measures changes in polarization as light interacts with an object and the resulting data are fitted with a `guess' model for the material (\emph{e.g.} a two-layer Cu$_2$O/Cu model or a three-layer CuO/Cu$_2$O/Cu one).

Computer simulations have also been widely applied to study the structural, optical and vibrational characteristics of copper oxides.
When carrying out simulations of materials at the atomic scale, classical empirical potentials (force fields) or more sophisticated quantum (\emph{ab initio}) approaches, such as density functional theory (DFT), can be used.
Force fields are parametrised empirical potentials tuned to reproduce the interactions of the atoms in the system at hand.
For certain problems force fields can provide a faithful description.
However, the reliability and transferability of such calculations depends primarily on the extent and quality of the data used in their construction.
Moreover, force fields cannot generally be applied to study chemical reactions.
Current development in force fields are addressing these issues.
Indeed, parametrisations obtained by fitting large data sets using \emph{e.g.}, neural network~\cite{neural_networks} and machine learning methods~\cite{szlachta_prb_2014}, to name just two, are improving accuracy and transferability.
Moreover, bond order potentials (such as ReaxFF~\cite{reaxff}) and polarisable force fields~\cite{warshel_1976} are making it possible to simulate chemical reactions.
It is difficult to make general statements about the sizes of systems that can be explored with various methods.
However, with many standard force fields it is now possible to examine systems with $\sim100,000$ atoms on a routine basis.
In addition, it is possible to explore the evolution of a system of this size, again on a routine basis, with an approach such as molecular dynamics for several hundreds of nanoseconds.

More accurate approaches are \emph{ab initio} methods which aim to study the structure and properties of a material by seeking (approximate) solutions to quantum mechanical equations such as the many-electron, many-atom Schroedinger equation.
These methods are more general and do not require system-specific parametrisations, however they are computationally more expensive.
One of the most widely used methods to study the properties of bulk materials and surfaces is density functional theory (DFT)~\cite{ks_pr_1965, hk_pr_1964, brazdova2013atomistic, curtarolo_natmat}.
With DFT it is possible to make genuine predictions about structural properties of a material within a few percent of the experimental value.
DFT also provides access to the electronic structure of the systems being considered and related spectroscopic properties.
In DFT, the energy of the electronic system is determined from the electronic density by solving Schroedinger-like equations.
Whilst exact in principle, in practice approximations have to be introduced since the functional form of the electron-electron interaction, called the exchange-correlation (XC) functional, is unknown.
Many approximate XC functionals have been developed~\cite{burke}, the most common ones being the local density approximation (LDA) and the generalised gradient approximation (GGA).
A number of deficiencies in DFT arise from these approximations and the choice of an appropriate XC functional is critical in order to obtain meaningful results.
As an example, in strongly correlated systems like CuO (typically, where $d$ and $f$ orbitals are localised), the GGA and LDA functionals provide a poor description of the electronic and bulk crystal structure.
Moreover, the band gap obtained using GGA or LDA in semiconductors or insulators is generally underestimated.
In these cases it is possible to add simple but somewhat \emph{ad hoc} corrections to the functionals (\emph{e.g.} the Hubbard U~\cite{himmetoglu}, self-interaction correction SIC~\cite{sic}) or to use more sophisticated hybrid functionals (\emph{e.g.} HSE06~\cite{hse}, PBE0~\cite{pbe0}) which incorporate some exact Hartree-Fock exchange.
Compared to standard empirical potential methods, DFT is much more computationally demanding, and, on a routine basis, systems with only a few 100 atoms can be examined and the dynamics of such systems explored for only a few tens of picoseconds.

Other computational methods which have been used in copper oxide simulations are the GW method~\cite{gw}, used when optoelectronic properties are of interest, since it is more accurate at predicting band structures than standard DFT XC functionals.
In addition there have been a number of Hartree-Fock~\cite{hf} studies and calculations with simpler approaches, such as the linear combination of atomic orbitals (LCAO) and tight binding methods~\cite{lcao}.

\section{Oxide structures}
\label{oxide_structures}

We are going to introduce here the bulk and surface structures of the two main copper oxides, Cu$_2$O and CuO.
Knowledge of the physical properties of these materials, and especially of the surface structures, is relevant background when trying to understand the formation and growth of the copper oxide.

The two most common forms of the oxide (shown in Fig.~\ref{fig:oxide_str}) are cuprite (or cuprous oxide, Cu$_2$O), the principal oxide at low temperatures and pressure, and tenorite (or cupric oxide, CuO), dominant at high temperatures and pressures~\cite{honjo}.
Another copper oxide structure, paramelaconite (Cu$_4$O$_3$), exists as a rare mineral found in hydrothermal deposits of copper.
Cuprite has long been known to be the primary oxide for copper at ambient conditions and there is considerable interest in its application to catalysis, optoelectronics and gas sensing and therefore a large amount of work has been done to determine its physical and chemical characteristics.
Tenorite has been studied less and still relatively little is understood about the structure of its surfaces, with only a handful of experimental and computational studies performed to that aim.

\subsection{Cu$_2$O bulk properties}
\label{cu2o_bulk}

\begin{figure}
\centering
\vspace{-60pt}
\includegraphics[width=130mm]{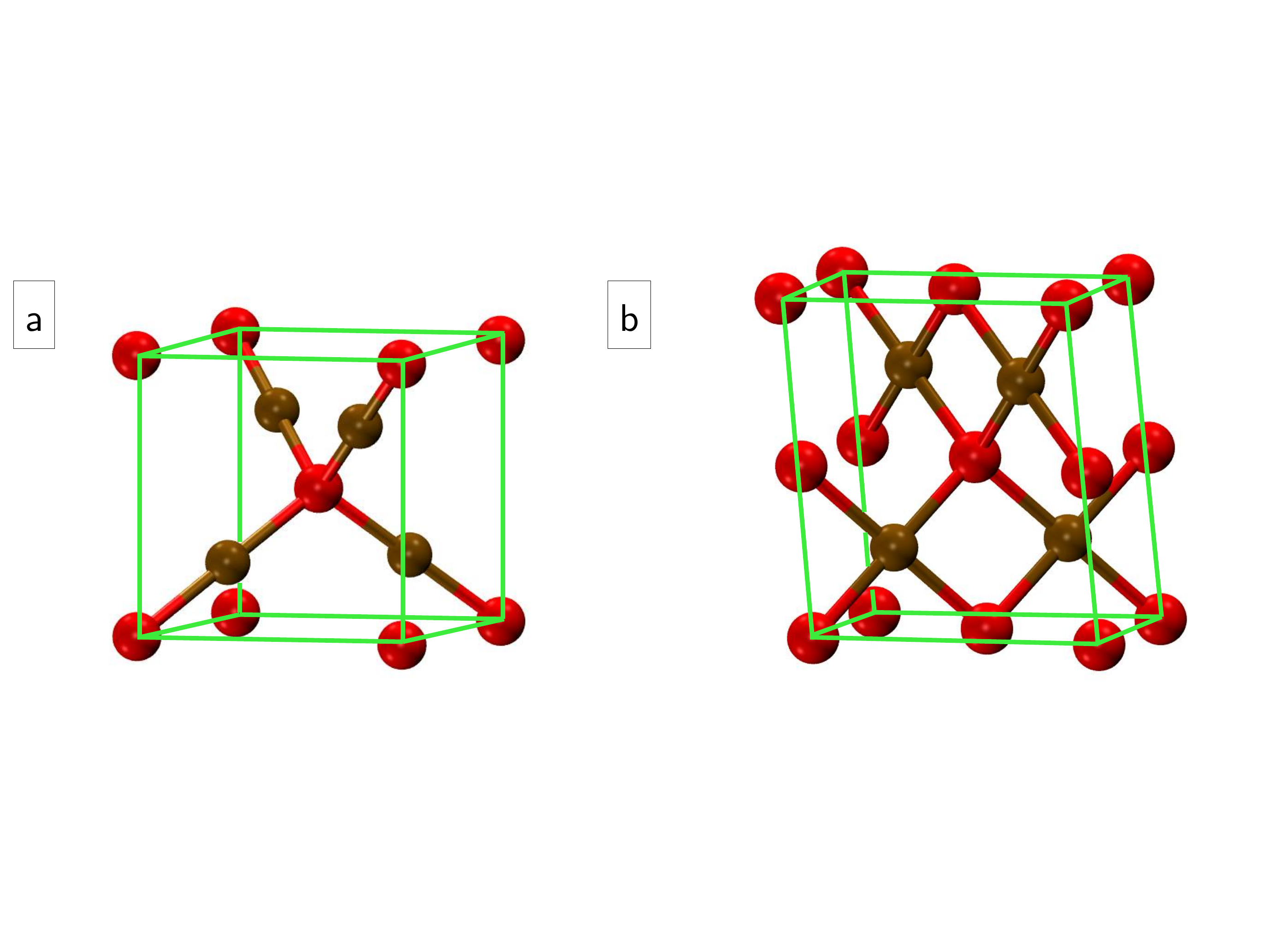}
\vspace{-50pt}
\caption{Ball-and-stick model of a unit cell of a) cuprous oxide Cu$_2$O and b) cupric oxide CuO. Red balls represent oxygen and brown balls copper. The unit cell is shown in green.}
\label{fig:oxide_str}
\end{figure}

In cuprous oxide (Cu$_2$O, see Fig.~\ref{fig:oxide_str}a), a cubic crystalline solid, copper has a Cu$^{1+}$ oxidation state. 
It is a $p$-type semiconductor with a direct band gap of $2.02-2.17$ eV and an optical gap of 2.62 eV~\cite{tandon, meyer, ghijsen_prb_1988}.
It is a promising material for a variety of industrial applications because of its band gap and because it shows negative thermal expansion~\cite{bohnen, mittal}.

The properties of cuprous oxide have been extensively studied using empirical potentials~\cite{hallil}, tight binding~\cite{robertson} and \emph{ab initio} methods.
\emph{Ab initio} studies include the use of DFT methods (see Table~\ref{table:cu2o}), and good agreement with the experimental bulk structure (\emph{i.e.} no more than $2.5\%$ discrepancy between the calculated and experimental value of the lattice constant) and vibrational modes~\cite{sanson} have been found.
However, the band gap is underestimated with standard DFT XC functionals, yielding values between $0.5$ and $0.8$ eV~\cite{islam_jms_2009, martinez_sss_2003, soon_prb_2006} and grossly overestimated with Hartree-Fock methods at 9.7 eV~\cite{ruiz_prb_1997}.
Corrections to DFT (\emph{e.g.} DFT+U or the GW approximation applied to DFT-GGA) give rise to band gaps that are in better agreement with experiment~\cite{raebiger,laskowski,filippetti_prb_2005, onsten,hu_prb_2008, schilfgaarde}, as summarised in Table~\ref{table:cu2o}.


\begin{table}
\begin{tabular}{|p{2.6cm}|p{2.5cm}|p{2cm}|p{2cm}|p{4.5cm}|}

\hline
Ref. & $a$ [\AA] & $B$ [GPa] & Band gap [eV]  & XC  \\ \hline 
Cortona~\cite{cortona_jpcm_2011} & 4.221 & 141 & -- & LDA  \\ \hline
Filippetti~\cite{filippetti_prb_2005} & 4.23 & -- & 0.55 & LDA  \\ \hline
Gordienko~\cite{gordienko_pss_2007} & 4.2696 & -- & 2.87 & LDA \\ \hline
Nie~\cite{nie} & 4.216 & -- & 0.52 & LDA  \\ \hline
Heinemann~\cite{heinemann2} & 4.1656 & -- & 0.99 & LDA+U (U-J=6.52)  \\ \hline
Tran~\cite{tran} & 4.27* & -- & 0.63-0.94 & LDA+U ($3<$U-J$<11$)  \\ \hline
Cortona~\cite{cortona_jpcm_2011} & 4.359 & 106 & -- & PBE \\ \hline
Islam~\cite{islam_jms_2009} & 4.312 & -- & 0.7 & PBE \\ \hline
Isseroff~\cite{isseroff_prb_2012} & 4.18 & 145 & 0.68 & LDA  \\ \hline
Isseroff~\cite{isseroff_prb_2012} & 4.10-4.17 & 135-143 & 0.81-1.15 & LDA+U ($2<$U-J$<8$) \\ \hline
Isseroff~\cite{isseroff_prb_2012} & 4.31 & 109 & 0.43 & GGA  \\ \hline
Isseroff~\cite{isseroff_prb_2012} & 4.26-4.30 & 96-106 & 0.54-0.84 & GGA+U ($2<$U-J$<8$) \\ \hline
Isseroff~\cite{isseroff_prb_2012} & 4.28 & 114 & 2.84 & PBE0  \\ \hline
Isseroff~\cite{isseroff_prb_2012} & 4.29 & 114 & 2.04 & HSE  \\ \hline
Le~\cite{le} & 4.317 & -- & -- & PBE  \\ \hline
Martinez-Ruiz~\cite{martinez_sss_2003} & 4.3 & 108 & 0.5 & PBE \\ \hline
Bohnen~\cite{bohnen} & 4.30 & 112 & -- & PBE  \\ \hline
Soon~\cite{soon_ss_2005} & 4.34 & 104 & -- & PBE \\ \hline
Soon~\cite{soon_prb_2006} & 4.32 & 104 & 0.64 & PBE  \\ \hline
Ruiz~\cite{ruiz_prb_1997} & 4.435 & 100 & 9.7 & HF  \\ \hline
Ruiz~\cite{ruiz_prb_1997} & 4.277 & 93 & -- & HF+LYP \\ \hline
Heinemann~\cite{heinemann2} & 4.2675 & -- & 2.02 & HSE06  \\ \hline
Scanlon~\cite{scanlon} & -- & -- & 2.12 & HSE  \\ \hline
Tran~\cite{tran} & 4.27* & -- & 0.79-2.77 & PBE0 \\ \hline
Bruneval~\cite{bruneval} & -- & --  & 1.34 & G$_0$W$_0$   \\ \hline
Bruneval~\cite{bruneval} & -- & --& 1.97 & scGW  \\ \hline
Filippetti~\cite{filippetti_prb_2005} & 4.23 & -- & 1.8 & SIC  \\ \hline
Kotani~\cite{kotani} & -- & -- & 1.97 & scGW  \\ \hline
Lany~\cite{lany} & -- & -- & 2.03 & GW+V$_d$ \\ \hline
\hline
Experiment~\cite{tandon, meyer} & 4.27 & 106-138 & 2.02-2.17 & -- \\ \hline

\end{tabular}
\caption{Summary of the calculated lattice constant $a$, bulk modulus $B$ and band gap for Cu$_2$O. Information on the exchange-correlation (XC) functional is also given. PBE is a type of GGA functional. scGW, G0W0 and GW+V$_d$ are different applications of the GW method. HF+LYP is a Hartree-Fock type calculation with an \emph{a posteriori} correction using the LYP correlation functional. For calculations using the +U Hubbard correction, values of the relevant parameters used (U-J) are given ion eV units. *These calculations were performed at the experimental value of the lattice constant.}
\label{table:cu2o}
\end{table}


\subsection{Cu$_2$O surfaces}
\label{cu2o_surf}

\begin{figure}
\centering
\vspace{-20pt}
\includegraphics[width=100mm]{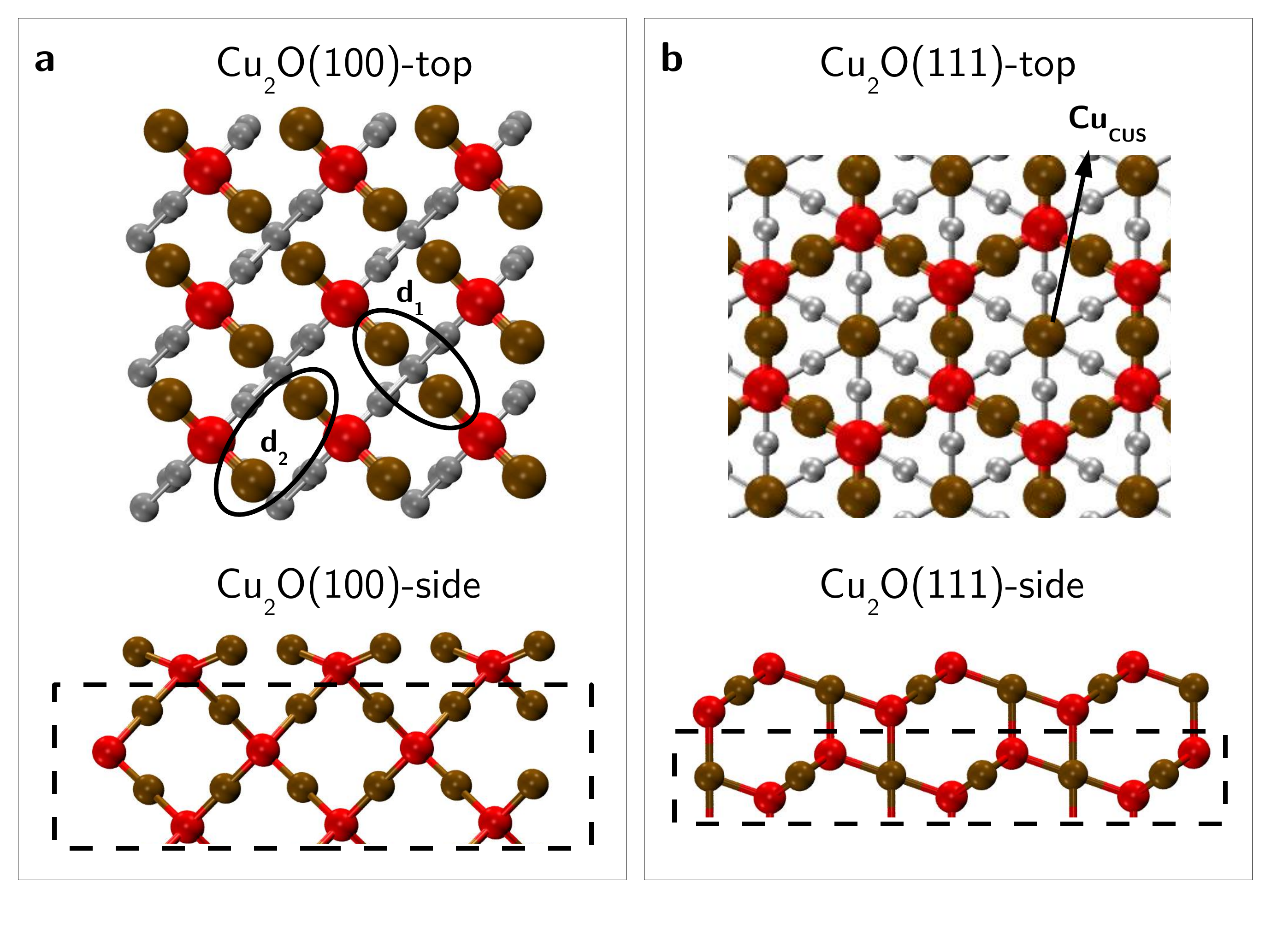}
\caption{Structure of the Cu-terminated Cu$_2$O(100) and O-terminated Cu$_2$O(111) surfaces. 
O atoms are shown in red and Cu atoms in brown.
Atoms below the surface layer are depicted in grey in the top views and enclosed in a box in the side views.
a) Top and side view of a copper terminated Cu$_2$O(100) slab, consisting of alternating layers of O and Cu atoms.
The surface copper atoms `sink' towards the O-atoms plane and form Cu-Cu dimers (indicated by d1). 
The second dimer formed in order to obtain the $(3\sqrt2 \times \sqrt2)R45^{\circ}$ reconstruction is labelled d2.
b) Top and side view of the O-terminated Cu$_2$O(111) surface. 
One O-Cu-O trilayer is shown between the dashed lines in the side view.
The surface layer presents a hexagonal structure with an unsaturated copper atom Cu$_\textrm{CUS}$ in the middle of each hexagon.}
\label{fig:soon}
\end{figure}

Experimental data are available only for Cu$_2$O(100) and Cu$_2$O(111) low index surfaces of cuprous oxide, by means of XPS (which monitors the Cu to O ratio on the surface), LEED in ultra high vacuum (which identifies the periodicity of the surface), and in the case of Cu$_2$O(111), also with TEM.

We first consider the polar Cu$_2$O(100) surface, which is formed by a succession of purely Cu or O layers and yields bulk-truncated surfaces which are either O- or Cu-terminated (Fig.~\ref{fig:soon}a).
Polar surfaces generally reconstruct~\cite{noguera_jpcd_2000} and three different reconstructions were observed on Cu$_2$O(100) according to the preparation method~\cite{schulz}.
The most stable surface in vacuum is a Cu-terminated surface with a $(3\sqrt2 \times \sqrt2)R45^{\circ}$ reconstruction.
The LEED pattern (which however showed many missing spots and the reconstruction is presented as tentative) was interpreted as a relaxation of the top layer of copper cations forming two surface dimers (d1 and d2 in Fig.~\ref{fig:soon}a).
Two transient structures were also identified: at 900 K, a $(\sqrt2 \times \sqrt2)R45^{\circ}$ reconstruction presenting a $1/2$ layer of terminal oxygen, and, after long oxygen exposures, the stoichiometric ($1 \times 1$) O-terminated surface.
Both structures revert to the $(3\sqrt2 \times \sqrt2)R45^{\circ}$ reconstruction upon further annealing over $500$ K.
The $(3\sqrt2 \times \sqrt2)R45^{\circ}$ reconstruction has been reproduced computationally~\cite{islam_ss_2010}, and it is more stable than the Cu- or O-terminated $(\sqrt2 \times \sqrt2)R45^{\circ}$ reconstruction~\cite{nygren, mcclenaghan, le}.

The other experimentally studied surface is Cu$_2$O(111).
A Cu$_2$O(111) slab is formed by a succession of O-Cu-O trilayers, ending with either a (stoichiometric, non polar) oxygen or (non-stoichiometric) copper termination (Fig.~\ref{fig:soon}b).
Photoemission and LEED experiments on this Cu-terminated (111) surface found, after annealing in UHV, a nearly stoichiometric reconstruction, with $(\sqrt3 \times \sqrt3)R30^{\circ}$ periodicity, attributed to an ordered 1/3 of an atomic layer of oxygen vacancies~\cite{schulz, onsten}.
Conversely, annealing in oxygen gives a stoichiometric oxygen-terminated surface (with possibly the loss of the unsaturated copper atoms, Cu$_{CUS}$ in Fig.~\ref{fig:soon}b, at the centre of the hexagons)~\cite{onsten}.
The stability of both polar and non-polar stoichiometric Cu$_2$O(111) surfaces has been studied also computationally\cite{islam_jms_2009,islam_ss_2009,li_ass_2011, li_pla_2010, soon_prb_2007, soon_ss_2007} by means of DFT combined with, in some cases, \emph{ab initio} thermodynamics.
\emph{Ab initio} thermodynamics (see \emph{e.g.}~\cite{reuter_2002, lozovoi_cpc_2001}) is a technique which allows one to estimate relative system stabilities in different environments; in this case, the stability of different oxide surfaces at a range of temperatures and oxygen partial pressures.
These studies showed that the experimentally observed Cu$_2$O(111) surfaces are indeed the most stable.~\cite{li_pla_2010}.

A wider range of low-index surfaces has been probed using DFT methods than experimentally~\cite{soon_prb_2007}.
The predicted lowest energy structure in high O$_2$ pressure is the already mentioned Cu$_2$O(111) surface missing the unsaturated Cu$_{CUS}$ atom and at low O$_2$ pressure the Cu$_2$O(110) surface with a CuO-like surface reconstruction, which has not been observed experimentally (the (110) surface has not been studied yet).

It is thus evident that there is scope for further experimental work on this topic, to confirm the computational predictions or to propose new reconstructions.
Surface-sensitive techniques such as STM and LEED, which allow a direct imaging of the surfaces, would be best suited to this aim.
Moreover, theoretical methods would be well suited to understand the reason behind the transitions between different reconstructions at different conditions of temperature and pressure.

\subsection{CuO bulk properties}
\label{cuo}

In CuO,  the copper atom has oxidation state Cu$^{2+}$.
The unit cell has monoclinic symmetry~\cite{asbrink} (see Fig.~\ref{fig:oxide_str}) and it contains four Cu$−$O dimers in the crystallographic unit cell, and two Cu$−$O units in the primitive cell. 
Each copper atom is located in the centre of an oxygen parallelogram. 
Each oxygen atom, in turn, has a distorted tetrahedral copper coordination.
CuO is a $p$-type semiconductor~\cite{ray_sem_2001,hardee_jes_1977,koffyberg,marabelli} and it is antiferromagnetic in its ground state~\cite{forsyth,yang_prb_1988,brown_jpcm_1991,ain_jpcm_1992}.

From a DFT point of view, standard XC functionals alone are not accurate enough to reproduce the distorted nature of the lattice, and, upon structural relaxation, they produce an orthorhombic rather than a monoclinic structure~\cite{peng_jap_2012}.
However, the use of more sophisticated functionals~\cite{debbichi, heinemann2, hu_jpcc_2010, peng_jap_2012, wu_prb_2006, laskowski} allows for the reproduction of the triclinic structure and good agreement with experiments on structural and vibrational data~\cite{REICHARDT,Dar20096279, chen_prb, kliche} (see Table~\ref{table:cuo}).


\begin{table}

\begin{tabular}{|p{2.2cm}|p{3.4cm}|p{1.0cm}|p{1.4cm}|p{2.2cm}|p{5.0cm}|}
\hline
Ref. &  lattice ($a$, $b$, $c$) [\AA] & $\beta$ [$^{\circ}$] & m$_B$ [$\mu_B$] & Band Gap [eV] & XC \\ \hline 
Peng~\cite{peng_jap_2012} &  4.05, 4.06, 5.06 & 90.02 & 0.0  & 0.0 & LSDA \\ \hline
Peng~\cite{peng_jap_2012} &  4.56, 3.27, 4.96 & 100.2 & 0.63 & 1.32 & LSDA+U (U=7.5, J=0) \\ \hline
Anisimov~\cite{anisimov_prb_1991} & -- & -- & 0.66 & 1.9 & LSDA+U (U=7.5, J=0.98) \\ \hline
Debbichi~\cite{debbichi} &  4.548, 3.305, 4.903 & 99.652 & -- & --  & LSDA+U (U=7.5, J=0.98) \\ \hline
Ekuma~\cite{ekuma}  &  4.68, 3.42, 5.13 & 90 & 0.68 & 1.25 & DFT+U (U=7.14) \\ \hline
Heinemann~\cite{heinemann2}  & 4.588, 3.354, 5.035 & 99.39 & 0.66 & 1.39 & LDA+U (U=7.5, J=0.98) \\ \hline
Wu~\cite{wu_prb_2006} & 4.55, 3.34, 4.99 & 99.507 & 0.6 & 1.0 & LSDA+U (U=7.5, J=0.98) \\ \hline
Hu~\cite{hu_jpcc_2010} &  --- & --- & 0.63 & 1.1  & GGA+S+U (U=7.5, J=0.98)\\ \hline
Jiang~\cite{jiang_jcp_2013} & 4.68, 3.42, 5.13 & 99.54 & 0.80 & --  & GGA+U (U=4.5)\\ \hline
Nolan~\cite{nolan} &  4.395, 3.846, 5.176 & -- & 0.53-0.7 & 0.17-2.11  & GGA+S+U (U=3-9, J=0)\\ \hline
Svane~\cite{svane} &  -- & -- & 0.65 & 1.43  & SIC-LSDA\\ \hline
Szotek~\cite{szotek} &  -- & -- & 0.64 & 1.0  & SIC-LSDA \\ \hline
Heinemann~\cite{heinemann2} &  4.513, 3.612, 5.141 & 97.06 & 0.54 & 2.74  & HSE06\\ \hline
Lany~\cite{lany} & -- & -- & -- & 1.19  & GW+V$_d$   \\ \hline
\hline
Exp.  & 4.684, 3.423, 5.129 \cite{asbrink} & 99.54 \cite{asbrink}& 0.68 & 1.2-1.9 \cite{meyer,ghijsen_prb_1988, koffyberg, pierson_ass_2003, marabelli} & --\\ \hline
\end{tabular}
\caption{Calculations of the CuO bulk structure (lattice parameters $a$, $b$, $c$ in \AA, angle $\beta$ (between the $a$ and $c$ axes) in degrees, magnetic moment $m_B$ in Bohr magnetons and bang gap in eV. The S in the functional name indicates spin-polarised calculations. All calculations featuring the Hubbard U (+U in the functional name) have been done with the so-called `Dudarev approach'~\cite{dudarev}, which requires two parameters, U and J. U and J are in eV.}
\label{table:cuo}
\end{table}


The research focus for CuO has been for a long time on its optical properties, for phototermal and photoconductive applications, and on magnetic and phase stability properties, for possible high-temperature superconducting applications.
The electronic properties of CuO, instead, have been studied only more recently.
The band gap of CuO has been experimentally measured to be $1.2-1.9$ eV~\cite{meyer,ghijsen_prb_1988, koffyberg, pierson_ass_2003, marabelli}, with the results depending on the sample preparation and on the measurement techniques used.
Computationally, the band gap is well reproduced with the LDA+U approach (see Table~\ref{table:cuo}), however the results depend strongly on the value of U while the HSE06 hybrid functional overestimates the gap by approximately 1 eV~\cite{heinemann2}.

\subsection{CuO surfaces}
\label{cuo}

The potential use of CuO nanostructures for catalysis, sensing or superhydrophobicity and many other applications~\cite{zhang_review} requires a good knowledge of the surfaces involved in these processes.
Only one experimental study has been performed to date on CuO surfaces.
In this work, it was found, using LEED, that the CuO(100) surface in UHV conditions does not present a reconstruction~\cite{warren_jpcm_1999}.

\begin{figure}
\centering
\vspace{-20pt}
\includegraphics[width=100mm]{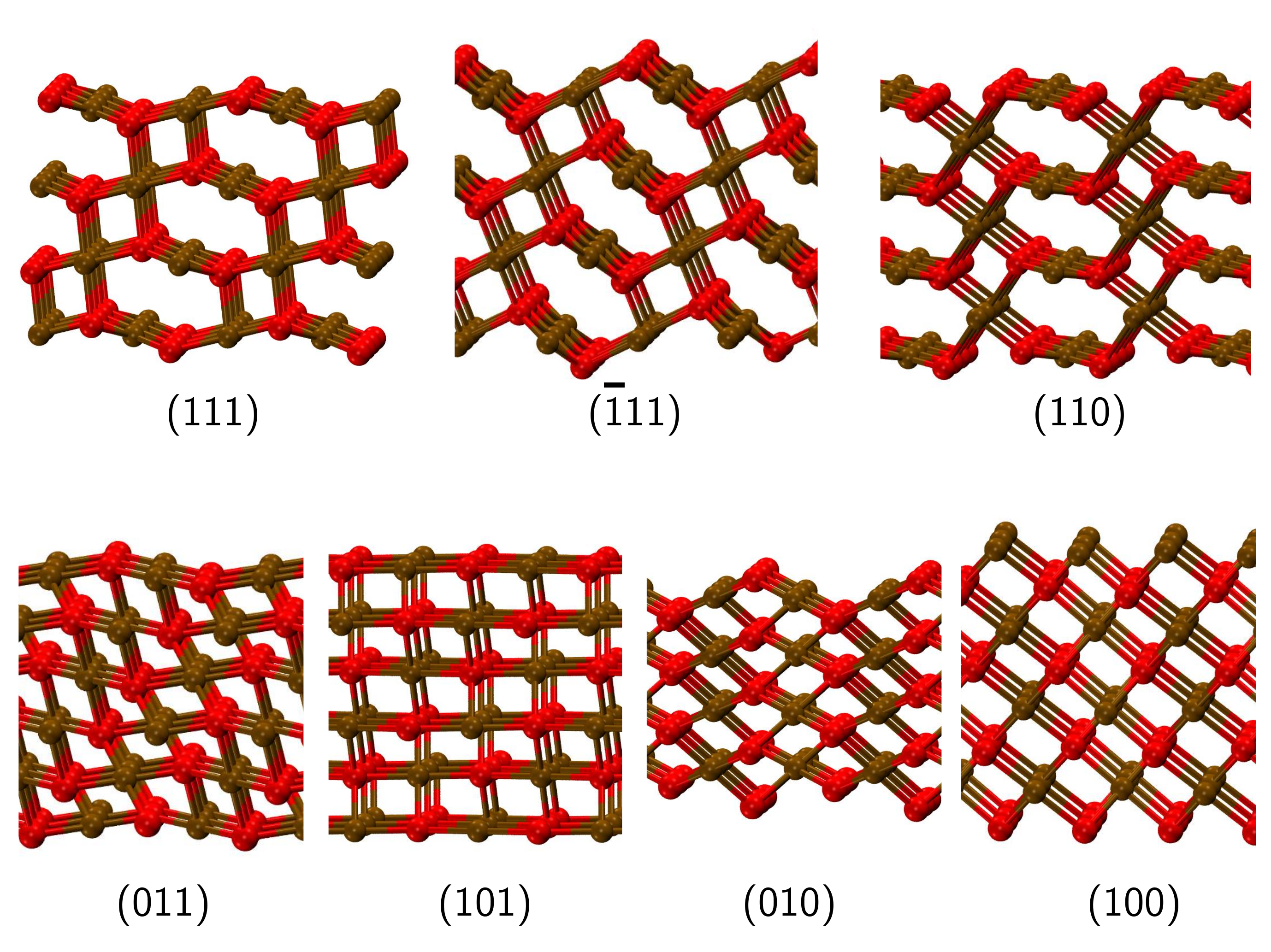}
\caption{Side views of CuO slabs showing the structure of stoichiometric cupric oxide CuO surfaces. They are shown in order of increasing surface energy (left to right, top to bottom), as calculated using \emph{ab initio} calculations, using the GGA+U method~\cite{hu_jpcc_2010}. Copper is shown in brown, oxygen in red.}
\vspace{-10pt}
\label{fig:cuo_surf}
\end{figure}

DFT studies (using a GGA XC functional with the Hubbard U correction combined, in some cases, with \emph{ab initio} thermodynamics) of the stoichiometric and of some defective surfaces have been performed~\cite{hu_jpcc_2010, maimaiti,Monte2014104}.
CuO(111) is the most stable surface also compared to defective CuO structures with surface and subsurface vacancies~\cite{maimaiti}, with surface energy $\gamma =0.74$ J/m$^2$ (although much higher than the lowest energy Cu$_2$O surface~\cite{soon_prb_2007}).
This is true for all oxygen partial pressures, except for a narrow range at very low O$_2$ pressures, where the Cu-terminated CuO(110) surface is favourable. 
The energy order of the stoichiometric surfaces is $(111)< (\overline{1}11) < (011) < (101) < (110) < (010) < (100)$ (shown in Fig.~\ref{fig:cuo_surf})~\cite{hu_jpcc_2010}.
The highest surface energies ($2.28$ J/m$^2$ for CuO(100)) are associated to large fields between anionic and cationic layers, such as for the (010) and (100) surfaces.
Among the non-stoichiometric surfaces,  the O- and Cu-terminated CuO(110) and CuO(100) are more stable than their stoichiometric counterpart for high (O-terminated) and low (Cu-terminated) oxygen pressures~\cite{hu_jpcc_2010}.

While important information on stoichiometric and defective CuO surfaces has been obtained using computational methods, experimental data are lacking and there is room for work to be performed in order to either confirm these predictions or to suggest more possible structures.

\section{Initial stages of oxidation: oxygen adsorption on clean copper surfaces}
\label{o_ads}

\begin{figure}
\centering
\includegraphics[width=100mm]{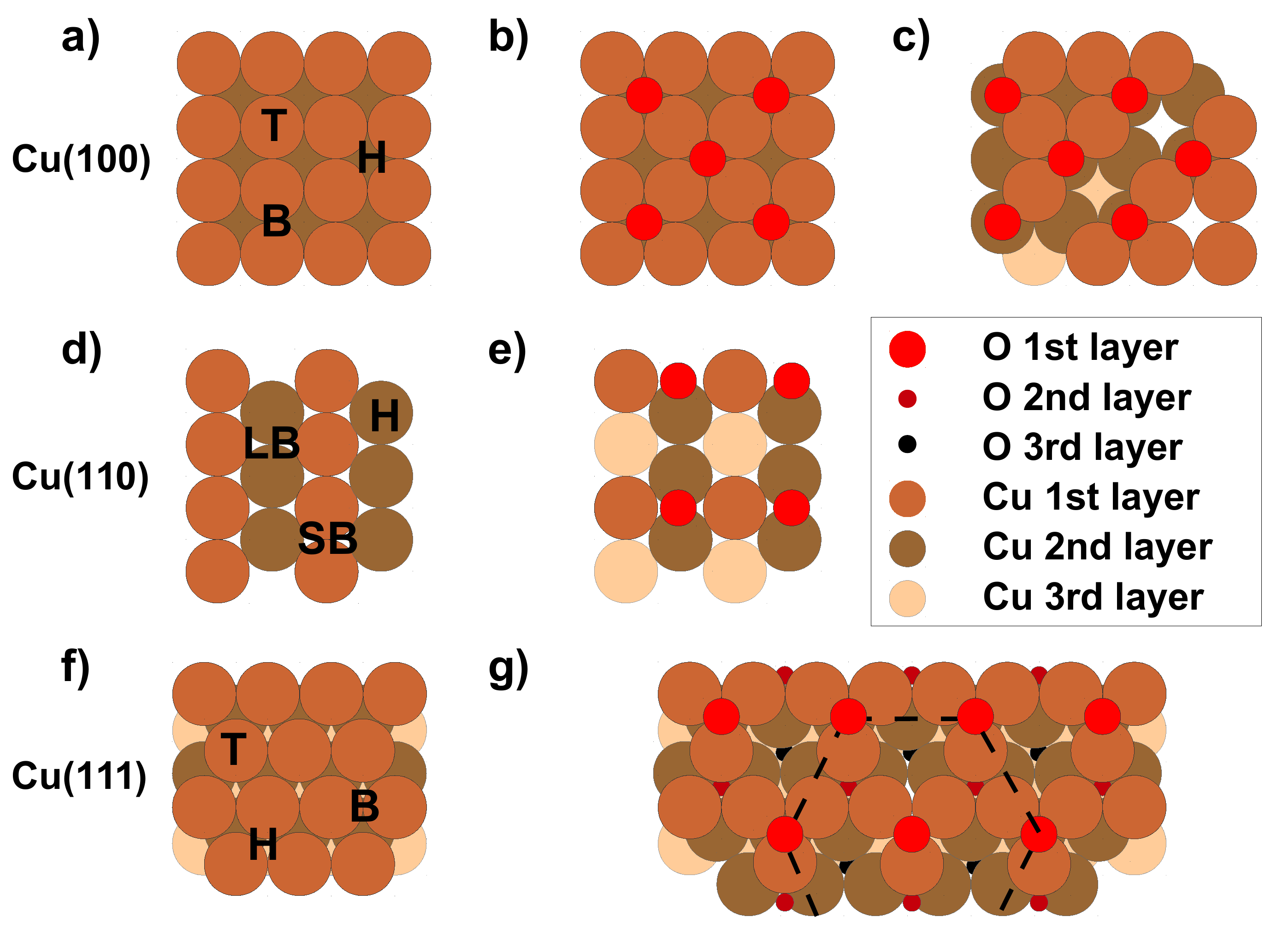}
\caption{Overview of the structures formed by oxygen on different copper surfaces, at low oxygen exposures. a, d, f) Clean surfaces of Cu(100), Cu(110) and Cu(111). T, H, B, SB and LB indicate respectively the top, hollow, bridge, short bridge and long bridge sites for adsorption. b) $c(2\times 2)$ overlayer structure seen on Cu(100) at coverages $< 0.3$ ML. c) $(2\sqrt 2 \times \sqrt 2)R45^{\circ}$ missing-row reconstruction seen on Cu(100) at coverages $> 0.5$ ML. e) $(2\times1)$ added-row reconstruction observed on the Cu(110) surface. g) Cu$_2$O(111)-like reconstruction of the Cu(111) surface having hexagonal geometry as shown by the dotted line. Two layers of subsurface oxygens are shown. Distorted variations of this reconstruction has been observed experimentally~\cite{jensen2, johnston_ss_2002}.}
\label{fig:cu_o_surf}
\end{figure}

The growth of an oxide can occur when a metal surface comes into contact with an oxygen-rich environment. 
The reaction sequence leading to oxidation of a clean metal surface is generally accepted to be oxygen chemisorption, nucleation and growth of the surface oxide, and bulk oxide growth.

It is well known that clean copper surfaces in vacuum do not reconstruct~\cite{crljen}.
However, exposure to oxygen pressure which is low enough not to trigger the formation of oxide-like structures induces reconstructions.
We will discuss the details of these reconstructions below, and whether they are important to the initial stages of oxidation is still openly debated.
It has been speculated that a reconstructed, O-saturated layer must form before the onset of oxidation, since dwell times, \emph{i.e.} the lapse of time between the beginning of oxygen deposition and observation of oxide formation, of up to 30 minutes have been observed~\cite{heinemann, dubois, yang_mm_1998}.
Moreover, evidence of the existence of the reconstructed copper surfaces and of subsurface growth of the oxide before the onset of island formation, has been produced using STM on Cu(100)~\cite{lahtonen} and Cu(111)~\cite{leon}.
DFT calculations have shown that, on Cu(100), subsurface oxide-like structures are more easily produced when the surface already has pre-adsorbed oxygen atoms, thus suggesting that the reconstruction facilitates oxide formation~\cite{lee_prb_2011}.
However, it has also been shown that upon exposure to oxygen at higher pressures~\cite{zhou_prl_2012} copper oxide formation at step sites occurs without prior surface reconstruction thus opening the debate about whether direct formation of oxide islands can occur without the formation of a O/Cu overlayer.

In the following section, the very initial stage of the oxidation process, \emph{i.e.} the chemisorption of oxygen onto copper surfaces is reviewed.
A summary of the overlayer structures formed on Cu after O-dosing is shown in Fig.~\ref{fig:cu_o_surf}.
Cu(100) presents two main reconstructions, one with $c(2\times 2)$ symmetry (Fig.~\ref{fig:cu_o_surf}b) and a missing-row reconstruction (Fig.~\ref{fig:cu_o_surf}c), which are discussed in depth in Sec.~\ref{ocu100}.
The main overlayer structure for Cu(110) is the added-row reconstruction shown in Fig.~\ref{fig:cu_o_surf}e and discussed in Sec.~\ref{ocu110}), while the O/Cu(111) system, presented in Sec.~\ref{ocu111} shows a distorted hexagonal structure resembling the Cu$_2$O(111) surface (Fig.~\ref{fig:cu_o_surf}g).

Oxygen adsorption on clean copper has been extensively studied experimentally and computationally, and it has been the subject of a number of reviews over the years, \emph{e.g.} see Refs.~\cite{rous, besenbacher}. 
Here we focus only on the studies which are relevant in order to understand the onset of copper oxidation on the Cu(100), (110) and (111) surfaces and on the most recent developments in the field.

\subsection{Cu(100)}
\label{ocu100}

\begin{figure}
\centering
\vspace{-30pt}
\includegraphics[width=120mm]{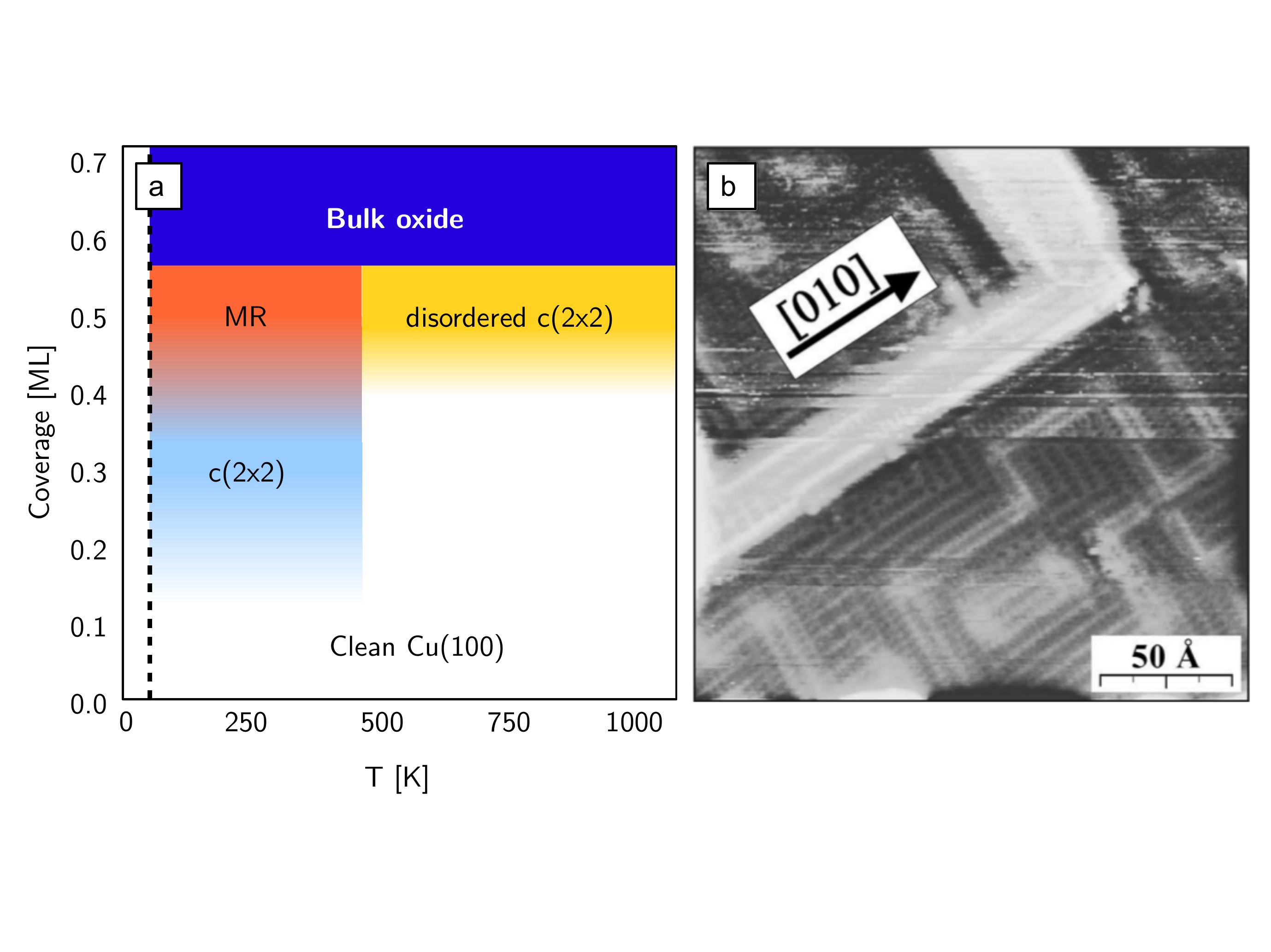}
\vspace{-50pt}
\caption{a) Summary of the structures forming on Cu(100) after oxygen exposure. The dashed line indicates the lowest temperature at which experiments have been performed ($30$ K) for this system. The areas in white represent domains where no ordered oxygen overlayers form. For low temperatures, at coverages of $\sim 0.3$ ML the $c(2\times2)$ structure forms and at $\sim 0.5$ ML the missing row (MR) reconstruction is seen. At intermediate coverages both structures exist. At high temperatures a `disordered' $c(2\times2)$ structure occurs. At coverages over $0.5$ ML subsurface oxide starts to grow. b) STM image of Cu(100) after O$_2$ exposure ($P=3.7\times10^{-2}$ mbar, $T=373$ K). A $(2\sqrt{2} \times \sqrt{2})R45^{\circ}$ island and a zigzag phase boundary (bright stripe) are visible. The missing rows of Cu run along the $\langle 001 \rangle$ directions and are imaged as depressions. Taken from Ref.~\cite{lahtonen}.}
\label{fig:cu_100lahtonen}
\end{figure}

Oxygen adsorption on Cu(100) has been widely studied~\cite{besenbacher, lahtonen, sun_srl_2001}, and an interesting variety of structures is formed depending on temperature and coverage, as shown in the phase diagram in Fig.~\ref{fig:cu_100lahtonen}a.

At low temperatures (up to $100$ K) and low coverage ($\sim 0.1$ monolayers, ML, defined as one adsorbed oxygen atom for every surface copper atom) experimental and computational evidence has shown that incident oxygen molecules dissociate with the oxygen atoms adsorbing at the hollow site~\cite{lederer_prb_1993, alatalo_prb_2004, chen_cjcp_2006, duan_prb_2010, yagyu_jpcc_2009, puisto_ct_2005}.
These dissociated oxygen atoms stabilize chemisorption of further incoming oxygen molecules at higher coverages~\cite{rajumon, spitzer_1982,yokoyama_prb_1993, katayama_jpcc_2007}.

Below $473$ K, two overlayer structures form: a $c(2\times 2)$ phase~\cite{lahtonen, fujita, arvanitis_cpl_1993, sotto_ss_1992, kittel_2001} at a coverage of $\sim 0.3$ ML and a $(2\sqrt 2 \times \sqrt 2)R45^{\circ}$ missing-row (MR) reconstruction~\cite{zeng, jensen_prb_1990,woll, leibsle, yata_1997, wuttig, wuttig_2,robinson_prb_1990,liu_ps_1995, bonini_ss_2006} at $\sim 0.5$ ML.
In the $c(2\times 2)$ phase (shown in Fig.~\ref{fig:cu_o_surf}b) the oxygen atoms occupy four-fold hollow sites on Cu(100) and form nanometre-sized $c(2\times 2)$ domains separated by oxygen-deficient zig-zag shaped boundaries. 
The MR reconstruction (Fig.~\ref{fig:cu_o_surf}c,~\ref{fig:cu_100lahtonen}b) can be viewed as a $c(2\times2)$ structure with each fourth [100] row of Cu atoms missing. 
Between 0.3 and 0.5 ML, Cu atoms are ejected from the $c(2\times2)$ domains~\cite{tanaka, lee_prb_2011} and MR islands start forming on terraces, until, at $0.5$ ML coverage, a network of missing-row reconstruction islands covers the whole surface~\cite{lahtonen, fujita}.

At high temperatures ($473<$ T $<1000$ K) a $c(2 \times 2)$-like state, with $25 \%$ vacancies in the top Cu layer forms at $0.5$ ML coverage, instead of the MR reconstruction~\cite{iddir_prb_2007}.

No further oxygen adsorption occurs above $0.5$ ML for experiments at very low pressures ( $ < \sim 10^{-7}$ Torr).
However, at higher pressures, further exposure of the surface to O$_2$ leads to growth of oxide in the subsurface regions, while the surface still exhibits the MR reconstruction~\cite{lahtonen, yata_1997}, validating the hypothesis that oxygen-induced surface reconstruction is indeed the first step of oxide growth.

DFT studies combined with \emph{ab initio} thermodynamics~\cite{duan_prb_2010,saidi_prb_2012} confirmed that the two experimentally observed reconstructions at lower temperatures are the most stable during the early stages of Cu(100) oxidation prior to the onset of bulk oxidation.

The transition between the two low-temperature reconstructions has been tentatively explained in terms of stress relief, electrostatics and orbital hybridization.
Compressive surface stress has been shown, both by means of experiments and DFT calculations~\cite{harrison, tanaka2} to increase with oxygen adsorption, and to be higher in the $c(2\times2)$ reconstruction.
Therefore the MR reconstruction is more stable at higher oxygen coverages.
Electrostatically, the driving mechanism for oxygen overlayer formation has been related to long-range Coulomb interaction~\cite{jaatinen_ss_2007, jaatinen_prb_2007,colbourn_prl_1991, stolbov_jcp_2002, stolbov_prl_2002,stolbov_prb_2002}, and the small size of the $c(2\times2)$ domains to repulsion between O adatoms and Cu and O adatoms. 
At coverages where the O atoms have nowhere to form distinct $c(2\times2)$ domains the phase transition occurs to minimize high O-O repulsion.
This is in disagreement with the findings of Merrick \emph{et al.}~\cite{merrick_2004}, who argue that the stability of the MR system is determined by orbital hybridization of neighbouring O-Cu which lowers the energy of the system, rather than by long-range interactions.

The transition between the ordered MR to the `disordered' $c(2\times2)$ reconstruction has been explained by means of DFT in terms of diffusion of Cu surface vacancies from an ordered array in the MR system to random positions as the temperature increases~\cite{iddir_prb_2007}.

An intermediate added row structure was predicted by Kangas \emph{et al.}~\cite{kangas_ss_2012} which has energies comparable to the MR reconstruction.
This reconstruction has however not yet been observed experimentally, possibly because the right conditions of temperature and pressure for this structure have not been used~\cite{lahtonen}.
\emph{Ab initio} thermodynamics calculations could approximately establish at which experimental conditions this reconstruction is expected and could thus inform further experimental work.

It is important to note that simulations have shown that the MR structure is a necessary step towards the formation of a Cu$_2$O-like structure~\cite{lee_prb_2011, lee_ss_2010}.
Indeed, Cu$_2$O-like structures were found to form on the MR reconstruction upon O adsorption which were not found on the non-reconstructed surface, thus confirming the experimental results of Lahtonen \emph{et al.}~\cite{lahtonen} and fitting well with the work of Zhou \emph{et al.}~\cite{zhou_cc_2013}, who finds the presence of a two-layer thick oxide before the formation of oxide islands of the Cu(100) surface (see Sec.~\ref{sec:100}).

In summary, three reconstructions can form on Cu(100) upon oxygen adsorption, as shown in Fig.~\ref{fig:cu_100lahtonen}.
Of these, the MR reconstruction is the most stable at room temperature, for coverages $\sim0.5$ ML.
Experimental and computational evidence has shown that the formation of this reconstruction is the first step towards the formation of the bulk oxide.

\subsection{Cu(110)}
\label{ocu110}

Molecular oxygen dissociates when deposited on Cu(110) at temperatures above $45$ K~\cite{sun_prb_2007}.
Upon dissociation of the O$_2$ molecules a number of overlayer structures are observed, including an added-row $(2\times1)$ structure (Fig.~\ref{fig:cu_o_surf}e) which is formed at an oxygen coverage of 0.5 ML~\cite{besenbacher, liu_ps_1995, feidenhansl_prb_1990,parkin_prb_1990,sun_prb_2004}. 
Another type of surface reconstruction, $c(6\times2)$, has also been reported at high coverage ($\sim 2/3$ ML)~\cite{ertl_1967, kishimoto_2008, feidenhansl, coulman_ss_1990, liu_ps_1995, dorenbos,sun_prb_2004, liu_ss_1995,liu_ss_2014} or at lower coverage but higher temperature~\cite{carley_ptrs_2005}. 

The $(2\times1)$ phase forms via the creation of Cu-O-Cu-O chains along the $[001]$ direction of the substrate, which eventually become the `added rows' on top of the clean substrate at a coverage of 0.5 ML.
The chains start forming above $70$ K but do not fully organize until $\sim 200$ K~\cite{sun_prb_2007}.
They are formed from mobile chemisorbed O atoms and Cu adatoms which leave from step edges and diffuse across the terraces~\cite{coulman_1990, jensen_prb_1990, chua_prl_1989, kuk_prb_1990, mocuta_1999, sun_prb_2007, li_cpl_2014}.
The experimental barrier calculated for the formation of these strings, $0.22 \pm 0.01$ eV~\cite{jacobsen_1990}, is close to the DFT-calculated barriers for Cu ($0.25$ eV) and O ($0.15$ eV) diffusion~\cite{liem_1998_ss}.

At oxygen coverages between 0.05 and 0.45 ML these Cu-O-Cu-O chains self-organize in a periodic array (called a supergrating) with a spacing varying between 60 and 140 \AA~\cite{kern_prl_1991, zhou_ss_2003,blanchard_2005, jensen_prb_1990, buisset_ss_1996}.
The dependence of the period of the supergrating as a function of oxygen coverage has been explained in terms of electrostatic~\cite{vanderbilt_1992} and elastic interactions~\cite{zeppenfeld_1994, prevot,berge_2003, bobrov_2008,harl_ss_2006, guillemot_prb_2011}.
In the first case, the period has been related to the difference in work function between the clean and reconstructed sections of the surface, in the second case to stress relief from the mismatch between the preferred period of the Cu-O-Cu-O chains and the period of the Cu(110) substrate in the [001] direction.
It could indeed be possible for both mechanisms to be at play.

The stability of the $(2 \times 1)$ reconstruction at 0.5 ML has been established computationally by means of semiempirical and DFT calculations, where it has been found to be energetically favourable with respect to the unreconstructed surface~\cite{jacobsen_1990, frechard_ss_1998, liem_1998_ss, liem_2000_ss} and to an alternative added row $(4 \times 1)$ geometry~\cite{harl_ss_2006}.
DFT simulations with \emph{ab initio} thermodynamics~\cite{duan_prb_2010, bamidele,liu_ss_2014} found that the Cu-O added row reconstruction is favoured at low oxygen exposures, whereas at higher oxygen exposures, a transition to the $c(6 \times 2)$ structure is predicted to occur.
Thus, experiment and theory are in qualitative agreement, although the absolute transition pressures vary enormously (by 10 orders of magnitude). 
Since it is a major challenge for current DFT XC functionals to accurately predict adsorption energies and the underlying value of the adsorption energy has a huge impact on subsequence pressure estimates, such a quantitive discrepancy is not uncommon~\cite{vdw_review, kresse_nat_mat, carrasco_nat_mat}.
The barrier to transition between the two structures has been calculated to be fairly high at $1.41$ eV, which seems to explain why this phase is observed only at high temperatures.

A large amount of work has been performed in order to understand the surface electronic states of the added row reconstruction, using both experimental methods such as photoemission spectroscopy~\cite{didio, cortona, courths, matzdorf} and theory~\cite{matzdorf, cabrera-sanfelix, pforte} .
The character of the O-Cu bonding is found to be predominantly ionic, and the surface O($2p$) orbitals hybridize strongly with the Cu($3d$) states, forming bonding and antibonding linear combinations, with the antibonding bands not having been identified unambiguously yet.

There is no evidence so far on how the observed reconstruction relates to the initial formation of the oxide, if at all.
DFT calculations of subsurface oxygen added beneath both reconstructed surfaces have found that when an O coverage of 1 ML is reached, subsurface oxide formation in the tetrahedral interstitial sites is predicted to occur~\cite{li_ss_2013}.
The presence of oxygen in the tetrahedral site has been linked to oxide formation (since the O in Cu$_2$O resides in the tetrahedral sites), and therefore this is a possible mechanism for the initial formation of the oxide, which should however be confirmed by experimental or further computational work.

\subsection{Cu(111)}
\label{ocu111}

The clean (111) surface has the lowest surface energy for copper and it is less reactive compared to the other low-index Cu surfaces. 
It is therefore less studied for oxygen adsorption~\cite{simmons, ho_1978, dubois, spitzer_1982, niehus_ss_1983,haase_ss_1988, rajumon,lu_jpcm_1991, jensen1, jensen2, vankooten_ss_1993, matsumoto_2001_ss,johnston_ss_2002, wiame_2007_ss}.
No ordered structures are observed experimentally for low oxygen exposure~\cite{niehus_ss_1983,dubois,haase_ss_1988, jacob_apa_1986}.
Indeed, also DFT studies have found that oxygen adsorbs preferentially at the threefold hollow site for coverages up to 0.75 ML, without forming periodic overlayer structures~\cite{mavrikakis_ss_2001,soon_prb_2006, ma_sc_2013, pang_lang_2007}.

At higher coverage,  the adsorbed oxygen and copper adatoms that are ejected from step edges and terraces~\cite{matsumoto_2001_ss, wiame_2007_ss, leon} start forming overlayer structures,  before the onset of epitaxial growth (see Sec.~\ref{sec:111}).
Two main classes of reconstructions have been proposed for the O/Cu(111) surface.
The first, seen in LEED studies, involves a Cu(100)-like overlayer, incommensurate with the underlying unreconstructed Cu(111) surface, with the oxygen atoms occupying the hollow sites~\cite{judd_ss_1986, toomes}.
The second class comprises two long-range ordered structures~\cite{jensen1, jensen2}, the  so-called `29' and `44' superstructures.
They have, respectively, ($\sqrt{13} R(46.1^{\circ}\times 7 R 21.8^{\circ}$) and ($\sqrt{73} R5.8^{\circ} \times \sqrt{21} R - 10.9^{\circ}$) symmetry and very large surface unit cells, 29 and 44 times larger than the $1\times1$ cell of clean Cu(111). 
They exhibit an honeycomb pattern formed by distorted honeycomb units of the `ideal' Cu$_2$O(111) overlayer shown in Fig.~\ref{fig:cu_o_surf}g.
The oxide-like overlayer structure has been confirmed by STM~\cite{matsumoto_2001_ss}, X-ray absorption studies~\cite{johnston_ss_2002} and DFT and \emph{ab initio} molecular dynamics~\cite{soon_ss_2007,soon_prb_2006}.
This differs from the behaviour of other O/transition metal systems such as O/Ag(111)~\cite{michaelides_jvsta_2005, schnadt_prl_2006} and O/Pd(111)~\cite{todorova_jpcb_2004}, where chemisorbed oxygen adlayers form, before the formation of a surface oxide.
Other systems, such as O/Ru(0001)~\cite{stampfl_ct_2005} and O/Rh(111)~\cite{gustafson_1, gustafson_2}, instead never form a surface oxide layer before the onset of bulk oxidation.

Therefore, both experimental and computational studies point to the formation of hexagonal or quasi-hexagonal structures on the Cu(111) surface upon oxygen adsorption, structures which can be viewed as the initial layer of a Cu$_2$O(111) film and which can potentially act as a template for the growth of further Cu$_2$O(111) layers.

\section{Oxide film growth}
\label{controlled}

When copper surfaces are exposed to a continuous flow of oxygen for a long time oxidation is expected to occur.
It is known in general, from well-defined surface science studies of metals, such as those discussed in Sec.~\ref{o_ads}, that the O/metal structure which forms at the onset of oxidation can be distinct from the bulk oxide surface structure~\cite{lawless, michaelides_cpl_2003, todorova_prb_2005}.
The growth of oxides in copper has been studied since the early 1920s and early work has been extensively reviewed (see \emph{e.g.} Ref.~\cite{lawless}).
In this section, we aim to present an overview of the main topics in the field of copper oxide growth mainly focusing on recent studies and on the present state of experiments and theory.

\begin{figure}
\centering
\vspace{-40pt}
\includegraphics[width=120mm]{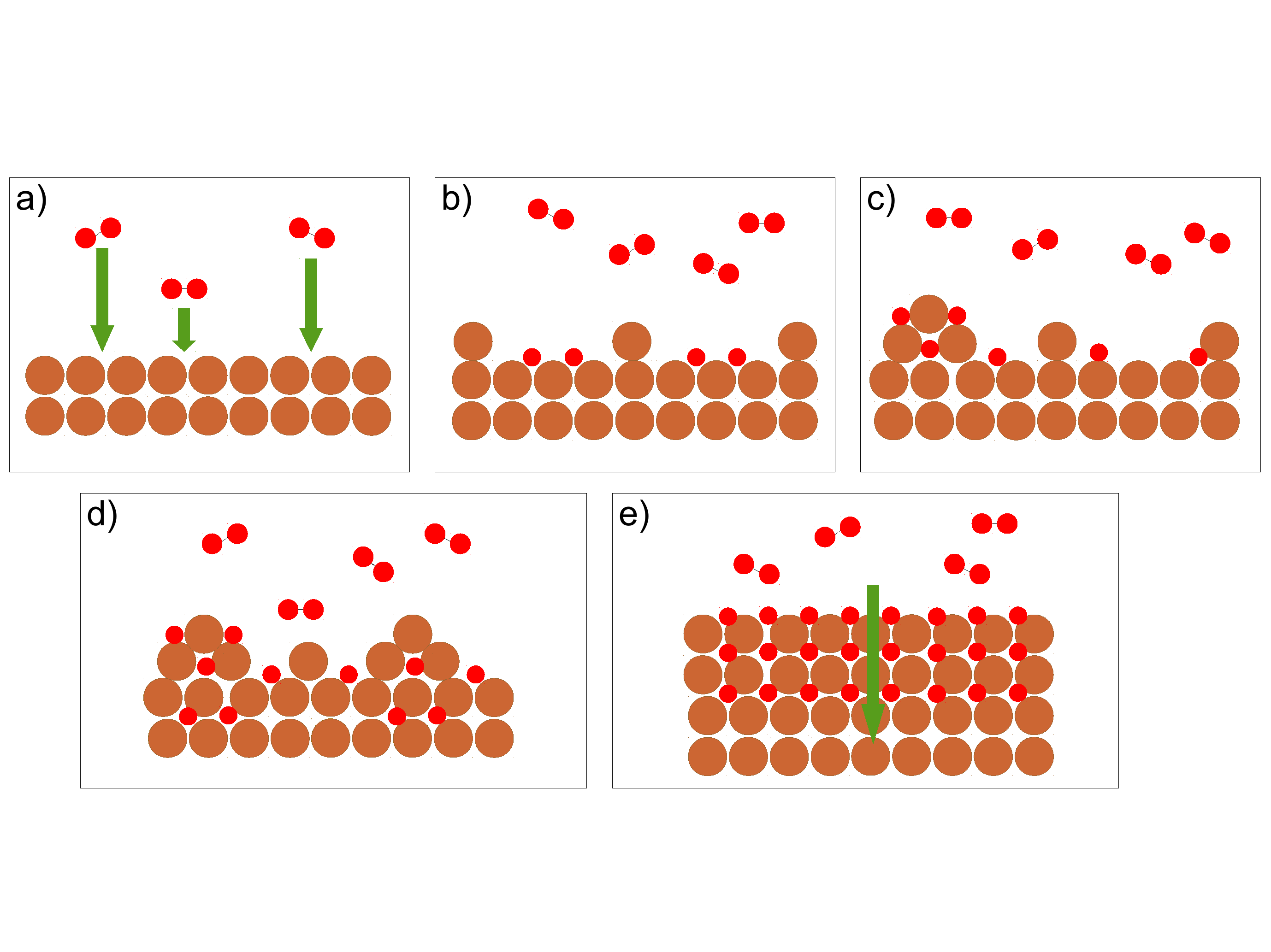}
\vspace{-45pt}
\caption{Schematic illustration of the stages of growth of a copper oxide film on copper. a) Gas-phase oxygen molecules and clean copper surface. b) Upon adsorption of the oxygen, the copper surface reconstructs. c) Nucleation of oxide islands upon diffusion of oxygen on the reconstructed surface, until the island saturation density ($N_{sat}$) is reached. d) Growth of the oxide islands. As the islands grow bigger direct oxygen impingement on the islands starts playing a more important role. e) After the islands coalesce oxide growth proceeds through interfacial diffusion of oxygen.}
\label{fig:cu_kinetics}
\end{figure}

In order to understand the microscopic details of the onset of oxidation, experiments have been performed in controlled laboratory conditions, where the orientation of the surface, temperature and oxygen pressures can be tuned to the required values.
The formation of the oxide has been directly observed by means of TEM, and their structure analysed by means of LEED and ellipsometry.
At the oxygen pressure, temperature and exposure times used in these studies, only cuprous oxide is expected to form~\cite{honjo,eastman, pierson}.
We review studies showing that the growth of cuprous oxide on low-index copper surfaces is epitaxial with the substrate~\cite{goulden, young, lawless, gwathmey,rhodin}, through nucleation and coalescence of nano-islands~\cite{brockway, milne_howie,simmons, dubois, goulden, heinemann, ho_1978} (as schematically represented in Fig.~\ref{fig:cu_kinetics}).
However, the mechanism of formation of the islands and the resulting shapes and growth rates are strongly dependent on the Cu substrate, as shown in Sec.~\ref{sec:100} for Cu(100), Sec.~\ref{sec:110} for Cu(110) and Sec.~\ref{sec:111} for Cu(111).
Key experimental findings are summarised in Tables~\ref{tab:uhv_100},~\ref{tab:uhv_110},~\ref{tab:uhv_111}.
The kinetic models for the nucleation and coalescence of the nano-islands and for the growth of the oxide thin film are reviewed in Sec.~\ref{controlled_4} and Sec.~\ref{controlled_1}.

Growth of the oxide in ambient conditions, when the copper surface is exposed to humid air, has also been extensively studied.
The body of work addressing this issue is reviewed in Sec.~\ref{controlled_1} and summarised in Table~\ref{table:native}.

Finally, the existence of preferential oxidation sites is discussed (Sec.~\ref{controlled_5}) and studies on the initial growth of CuO are reviewed (Sec.~\ref{controlled_cuo}).

\subsection{Oxide nano-islands: Cu(100)}
\label{sec:100}

\begin{figure}
\centering
\includegraphics[width=130mm]{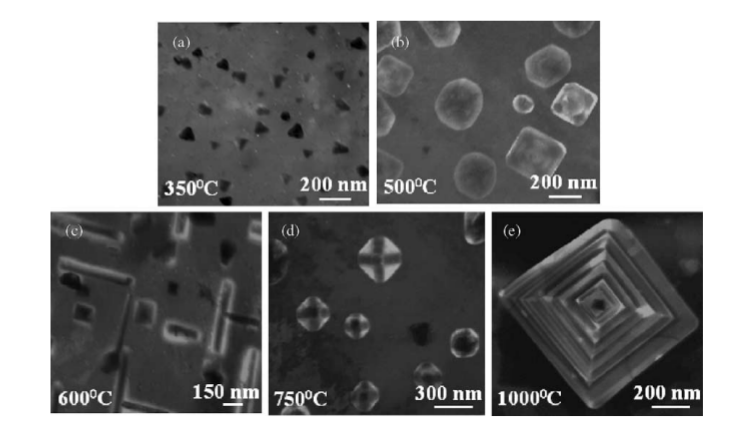}
\caption{Oxide islands observed on the Cu(100) surface using \emph{in situ} TEM imaging, for a set of temperatures between $350-1000$ $^{\circ}$C. The oxidation partial pressure is $P=5 \times 10^{-4}$ Tor. A large variety of island structures are observed, according to the oxidation temperature. Taken from Ref.~\cite{zhou_ass_2003}.}
\label{fig:cu100_islands}
\end{figure}

As for the formation of O/Cu overlayers, Cu(100) is the most extensively studied surface, both experimentally and computationally.
Copper oxidation on this surface proceeds, for low oxygen partial pressures ($P \sim 10^{-4}$ Torr), through the formation of islands which are epitaxial 
with the surface, {\it i.e.} Cu$_2$O(100) $||$ Cu(100)~\cite{yang_apl_1997,yang_SM_1998, yang_mm_1998, yang_apl_2002, brockway, heinemann, stefanov} and have $6 \times 7$ lattice misfit configuration~\cite{zhou_am_2009, eastman}.
These islands grow and coalesce with further oxygen deposition.
The shape of the islands depends on temperature~\cite{zhou_ass_2003} in a rather interesting manner as shown in Fig.~\ref{fig:cu100_islands}.
Below $350$ $^{\circ}$C only triangular islands are observed, whereas between $400-550$ $^{\circ}$C the shape changes to round or square.
At around $600$ $^{\circ}$C the islands start to grow until, at a critical size of $\sim 110$ nm, when they switch to a quasi-one-dimensional elongated rod shape.
Between $650-800$ $^{\circ}$C pyramid-shaped islands are observed and between $800-1000$ $^{\circ}$C hollow pyramids form~\cite{zhou_prl_2005}.
The big effect of temperature on the morphology of the islands could be either due to the dependence of copper and oxygen diffusion on temperature or to changes in interfacial strains and in the mechanical properties.
Indeed, the transition between the round/square to rod shape shown in Fig.~\ref{fig:cu100_islands}b,c has been explained as a competition between the surface energy and the strain of the islands due to the mismatch between the clean metal and the oxide lattices, following the so-called Tersoff-Tromp~\cite{tersoff_tromp} energy model.

TEM, XPS and AES studies looking at the cross section of an oxidising Cu(100) single crystal under low oxygen partial pressure~\cite{zhou_cc_2013, lampimaki_jcp_2007, yang_mm_1998, yang_mm_2001} have revealed that the oxide islands form on top of an oxide wetting layer.
The wetting layer itself has a $(\sqrt 2 \times 2 \sqrt 2) R 45^{\circ}$ missing row reconstruction and forms from the Cu atoms ejected by the missing row reconstruction of the substrate (described in Sec.~\ref{ocu100}) and the further oxygen deposited on the substrate~\cite{lampimaki_jcp_2007}.
Island growth also occurs beneath the surface, in good agreement with the already reviewed results by Lahtonen \emph{et al.}~\cite{lahtonen} (Sec.~\ref{ocu100}) who observed subsurface oxide growth for oxygen coverages above 0.5 ML and pressures above $\sim 10^{-7}$ Torr.

In general, when a metal is capable of forming uniform subsurface oxides (as in the case of Ag(110)~\cite{bao_cl_1995} and Ru(0001)~\cite{bottcher_jpc_1999}), oxide growth proceeds uniformly, rather than via island formation.
The reason for the island formation on top of the wetting layer is found in stress mismatch between the Cu(100) substrate and the Cu$_2$O(100) film~\cite{zhou_cc_2013}.

The transition between the O/Cu(100) system described in Sec.~\ref{ocu100} and the oxide islands has been studied to some extent with DFT, and although the full transition has not been modelled yet, insight has been gained into the mechanisms at play.
On the MR reconstructed surface, sub-surface adsorption becomes favourable for O coverage above $1.0$ ML, with low barriers for the transport of oxygen atoms below the surface~\cite{kangas_ss_2008, kangas_ss_2005, lee_ss_2009}.
Sub-surface oxygen atoms below the missing-row reconstruction adsorb in the tetrahedral interstitial sites, and thus form an oxide-like structure, unlike the case of sub-surface oxygen atoms below non-reconstructed surfaces, which adsorbs in the octahedral site~\cite{lee_ss_2010}.
However, the limiting factor for the oxidation of Cu(100) is the dissociation and on-surface diffusion of the oxygen molecule.
The dissociation of the oxygen molecule, although almost barrierless on the clean Cu(100)~\cite{alatalo_prb_2004}, is blocked by the on-surface oxygen on the reconstructed surface~\cite{alatalo_ss_2006}, and requires the diffusion of the oxygen molecule towards either a high-Cu concentration area~\cite{jaatinen_prb_2007} or vacancies and surface defects~\cite{ahonen_cpl_2008, junell_srl_2004} where the dissociation barrier is lower.
Diffusion of O and Cu atoms on the MR reconstructed surface is slow, having barriers of $1.4$ eV for oxygen and $2.0$ eV for Cu~\cite{jaatinen_prb_2007}.
The high barriers and slow diffusion make these processes difficult to simulate.
Indeed, Devine \emph{et al.}~\cite{devine_prb_2011} successfully simulated oxygen molecule dissociation on the clean Cu(100) surface with bond-order potentials, however they did not see any dissociation event on the missing-row reconstructed surface within their (short) 10 ps of molecular dynamics.
This highlights the difficulty of reproducing these complex phenomena which include diffusion of oxygen molecules, dissociation, diffusion of oxygen and copper atoms on and through the surface using standard computational approaches.
To this end, the combination of an accurate reactive potential with methods which allow for rare events to be probed, such as metadynamics~\cite{metadynamics} are possibly a way to go for this type of system.

{\footnotesize
\begin{longtable}{|p{1.8cm}|p{1.5cm}|p{4cm}|p{2.5cm}|p{1.5cm}|p{3.5cm}|}
\hline
{\bf Reference} & {\bf Sample thickness} & {\bf Surface Preparation} & {\bf Exp. Conditions} & {\bf Technique} &{\bf Result} \\ \hline
Brockway~\cite{brockway} (1972) & $90$ nm & Annealing in H$_2$ at T=$630$ $^{\circ}$C & P=$10^{-3}$ Torr, T=$525$ $^{\circ}$C & TEM & Epitaxial growth of oxide islands. \\ \hline
Heinemann~\cite{heinemann} (1975) & $80$ nm & Annealing, argon ion sputter etching & P=$5 \times 10^{-3}$ Torr, T=$425$ $^{\circ}$C & TEM & Epitaxial growth of square and hexagonal oxide islands. \\ \hline
Stefanov~\cite{stefanov} (1988) & -- & Ion bombardment and heating under UHV & $10$ to $2\times 10^{6}$ L, T=$-130-180$ $^{\circ}$C  & HREELS, XPS & Cu$_2$O formation at oxygen exposures between $10^{5}-2\times10^{6}$ L \\ \hline
Yang~\cite{yang_apl_1997} (1997) & $40$ nm  & Annealing in methanol vapour at $350$ $^{\circ}$C & $1.5 \times 10^{-5}$ Torr & TEM & Rate of growth of the oxide islands $\propto t^{1.3}$. \\ \hline
Yang~\cite{yang_mm_1998} (1998) & $100$nm & Annealing in UHV/CH$_4$O vapour at $350$ $^{\circ}$C & $1 \times 10^{-5}-1 \times 10^{-4}$ Torr & TEM & O monolayer forms before growth of the oxide islands.\\ \hline
Yang~\cite{yang_apl_1998} (1998) & $60$ nm  & Annealing in methanol vapour at $350$ $^{\circ}$C, $5 \times 10^{-5}$ Torr & $5 \times 10^{-5}-760$ Tor at $60-600$ $^{\circ}$C & TEM & Epitaxial island formation after surface reconstruction. \\ \hline
Yang~\cite{yang_SM_1998} (1998) & $100$ nm  & Annealing in CH$_4$O vapour at $350$ $^{\circ}$C, $5 \times 10^{-5}$ Torr & $5 \times 10^{-4}$ Tor $290-435$ $^{\circ}$C & TEM & Nucleation of islands promoted by O diffusion.\\ \hline
Yang~\cite{yang_jes_1999} (1999) & $100$ nm  & Annealing in CH$_4$O vapour at $350$ $^{\circ}$C, $5 \times 10^{-5}$ Torr & $1 \times 5^{-5}-5 \times 10^{-4}$ Tor $70-600$ $^{\circ}$C & TEM & Preferential nucleation site at the edge of holes. \\ \hline
Yang~\cite{yang_mm_2001} (2001) & $60-100$ nm  & Annealing in CH$_4$O vapour at $350$ $^{\circ}$C, $5 \times 10^{-5}$ Torr & O$_2$/H$_2$O vapour at $5 \times 10^{-4}$ Tor at $350$ $^{\circ}$C & TEM & Initial surface reconstruction prior to island growth. \\ \hline
Yang~\cite{yang_apl_2002} (2002) & $60$ nm & Annealing in CH$_4$O vapour at $350$ $^{\circ}$C, $5 \times 10^{-5}$ Torr & $5 \times 10^{-5}-760$ Torr at $60-600$ $^{\circ}$C & TEM & Good agreement of kinetic data with the JMAK model. \\ \hline
Zhou~\cite{zhou_ass_2003} (2003) & $70$ nm & Annealing in CH$_4$O vapour at $350$ $^{\circ}$C, $5 \times 10^{-5}$ & $5 \times 10^{-4}$ Torr at $150-1000$ $^{\circ}$C & TEM & Temperature-dependent shape of oxide nano-islands. \\ \hline
Eastman~\cite{eastman} (2005) & $110-200$ nm & Annealing in Ar-$2\%$H$_2$ at $850$ $^{\circ}$C, $5 \times 10^{-5}$ Torr & $5 \times 10^{-4}$ Torr at $350-780$ $^{\circ}$C & X-ray scattering & Epitaxial nano-island formation.\\ \hline
Zhou~\cite{zhou_jap_2005} (2005) & $70-80$ nm  & Annealing in CH$_4$O vapour at $350$ $^{\circ}$C/vacuum $800^{\circ}$C & $5 \times 10^{-4}$ Torr, $350$ $^{\circ}$C & TEM & Island nucleation rate faster on (111) than (110) or (100) \\ \hline
Zhou~\cite{zhou_jmr_2005} (2005) & $70-80$ nm  & Annealing in vacuum at $550$ $^{\circ}$C & $5 \times 10^{-5}$ Torr, $350-900$ $^{\circ}$C & TEM & Temperature-dependent shape and oxidation rate of the oxide islands. \\ \hline
Zhou~\cite{zhou_prl_2005} (2005) & $70-80$ nm  & Annealing in CH$_4$O vapour at $350$ $^{\circ}$C/vacuum $800$ $^{\circ}$C & $5 \times 10^{-4}$ Torr, $350$ $^{\circ}$C & TEM & Island nucleation rate faster on (111) than (110) or (100) \\ \hline
Lampimaki~\cite{lampimaki_jcp_2007} & -- & Ar$^+$ bombardment, annealing at $700$ $^{\circ}$C & $2.8 \times 10^{-2} - 160$ Torr, $30-100$ $^{\circ}$C & XPS, XAS, STM & Island formation on top of the missing-row reconstruction for O/Cu(100) \\ \hline
Lahtonen~\cite{lahtonen} (2008) & $0.5$ mm & Ar$^+$ bombardment, annealing at $430$ $^{\circ}$C & $6\times10^{-7}$ Torr and $2.8\times10^{-2}$ Torr, $T=100$ $^{\circ}$C & STM & Surface reconstruction and oxide island formation at high O$_2$ exposure. \\ \hline 
Zhou~\cite{zhou_am_2009} (2009) & $70$ nm  & Annealing in Ar/H$_2$ at $700$ $^{\circ}$C & $5 \times 10^{-5}$ Tor, $700$ $^{\circ}$C & TEM & Cu$_2$O islands, $200/500$ nm side. Epitaxial growth with inclined Cu$_2$O/Cu edges. \\ \hline
Zhou~\cite{zhou_prl_2012} (2012) & $50$ nm  & Annealing in H$_2$ at $600$ $^{\circ}$C & $5 \times 10^{-5}$ Tor, $350$ $^{\circ}$C & TEM & Step-edge induced oxide growth. \\ \hline
Zhou~\cite{zhou_cc_2013} (2013) & $50$ nm  & Annealing in H$_2$ at $600$ $^{\circ}$C & $1 \times 10^{-3}$ Tor, $550$ $^{\circ}$C & TEM & Cu$_2$O islands grow on an oxide wetting layer nucleated on surface steps. \\ \hline
\caption{Experimental results of copper oxide formation in ultra-high vacuum on Cu(100). Experimental conditions and techniques have been listed, together with a summary of the main result of each study.}
\label{tab:uhv_100} 
\end{longtable}}

\subsection{Oxide nano-islands: Cu(110)}
\label{sec:110}
The islands formed on Cu(110) single crystals after being exposed to low pressure oxygen ($P \sim 10^{-4}$ Torr) are also found to be epitaxial with the substrate~\cite{zhou_ss_2003}. 
The island morphology and the time required to reach saturation density depend on temperature, with higher temperatures requiring much shorter time to saturation.
Between $450$ and $650$ $^{\circ}$C the lateral size of the islands is almost constant ($200-250$ nm), however the islands were found to thicken considerably ($24-40$ nm) beneath the Cu surface, showing that the rise in temperature greatly enhances the interfacial diffusion of oxygen~\cite{zhou_ass_2004}.
At $700$ $^{\circ}$C, the (110) surface of clean copper roughens with a step height of $\sim 10$ nm~\cite{zhou_ss_2004}.
Exposure to oxygen of this rough surface shows the formation of a higher density of oxide islands with a fast nucleation rate but slower lateral growth than for smooth surfaces.
This comparison between smooth and rough surfaces supports the idea that the kinetics of nucleation and growth of islands is dependent on the surface diffusion of oxygen atoms on
the surface (see Sec.~\ref{controlled_4}).

{\footnotesize
\begin{longtable}{|p{1.8cm}|p{1.5cm}|p{4cm}|p{2.5cm}|p{1.5cm}|p{3.5cm}|}
\hline
{\bf Reference} & {\bf Sample thickness} & {\bf Surface Preparation} & {\bf Exp. Conditions} & {\bf Technique} &{\bf Result} \\ \hline
Zhou~\cite{zhou_prl_2002} (2002) & $70$ nm & Annealing in CH$_4$O vapour at $350$ $^{\circ}$C, $5 \times 10-5$ Tor & $8/1 \times 10^{-4}$ Tor at $600$ $^{\circ}$C & TEM & Rod-shaped $20$ nm-thick islands.\\ \hline
Zhou~\cite{zhou_ss_2003} (2003) & $70$ nm & Annealing in CH$_4$O vapour at $350$ $^{\circ}$C & $5 \times 10^{-4}$ Tor, $300$ $^{\circ}$C and $450$ $^{\circ}$C & TEM & Faster initial oxidation rate than Cu(100) \\ \hline
Zhou~\cite{zhou_ss_2004} (2004) & $70$ nm & Annealing in CH$_4$O vapour at $350$ $^{\circ}$C & $5 \times 10^{-4}$ Tor, $350$ $^{\circ}$C and $750$ $^{\circ}$C & TEM & Higher density and slower lateral growth rate of islands on rougher Cu(110) surfaces.\\ \hline
Zhou~\cite{zhou_ass_2004} (2004) & $70$ nm & Annealing in CH$_4$O vapour at $350$ $^{\circ}$C & $5 \times 10^{-4}$ Tor, $750$ $^{\circ}$C & TEM & Faster oxidation at higher temperatures.\\ \hline
Zhou~\cite{zhou_jap_2005} (2005) & $70-80$ nm & Annealing in CH$_4$O vapour at $350$ $^{\circ}$C/vacuum $800^{\circ}$C & $5 \times 10^{-4}$ Tor, $350$ $^{\circ}$C & TEM & Island nucleation rate faster on (111) than (110) or (100) \\ \hline
Zhou~\cite{zhou_jmr_2005} (2005) & $70-80$ nm & Annealing in vacuum at $550$ $^{\circ}$C & $5 \times 10^{-5}$ Tor, $350-900$ $^{\circ}$C & TEM & Temperature-dependent shape and oxidation rate of the oxide islands. \\ \hline
Zhou~\cite{zhou_prl_2005} (2005) & $70-80$ nm & Annealing in CH$_4$O vapour at $350$ $^{\circ}$C/vacuum $800$ $^{\circ}$C & $5 \times 10^{-4}$ Tor, $350$ $^{\circ}$C & TEM & Island nucleation rate faster on (111) than (110) or (100) \\ \hline
\caption{Experimental results of copper oxide formation in ultra-high vacuum on Cu(110). Experimental conditions and techniques have been listed, together with a summary of the main result of each study.}
\label{tab:uhv_110} 
\end{longtable}}

\subsection{Oxide nano-islands: Cu(111)}
\label{sec:111}
\begin{figure}[h]
\centering
\includegraphics[width=100mm]{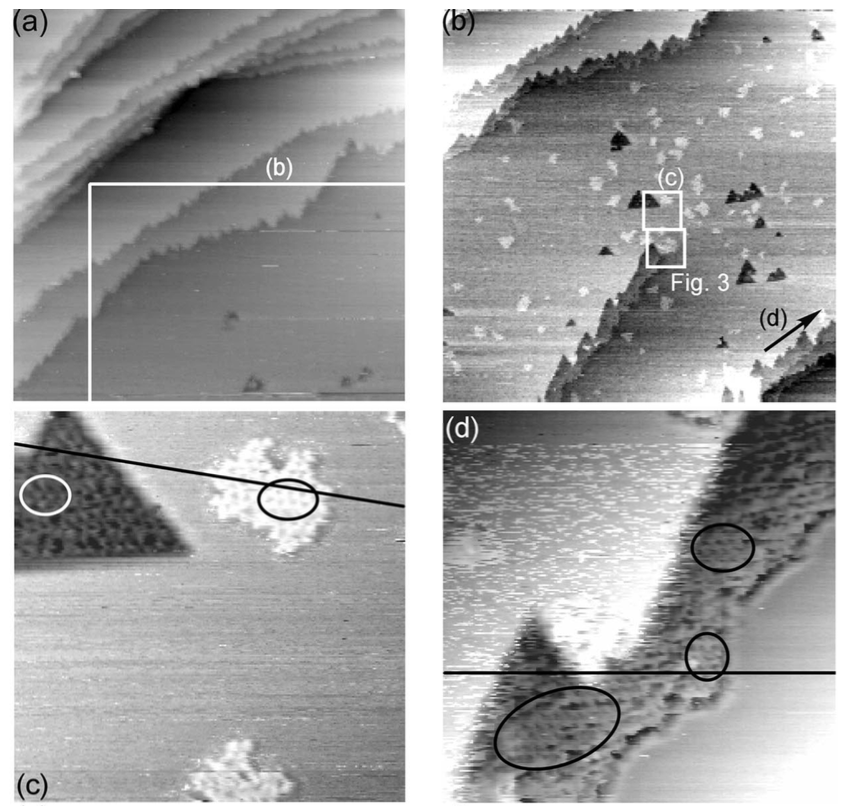}
\caption{Structures observed with STM by Matsumoto \emph{et al.}~\cite{matsumoto_2001_ss} when oxidising a Cu(111) surface with step edges (visible in panel a). In-terrace and on-terrace oxides are visible in panel b, and atomically resolved in panel c. In-terrace oxide is darker and on-terrace oxide lighter than the clean copper surface. Panel d shows the oxide which grows at the edge of the terrace. The area show in panel a,b is $2000 \times 2000$ \AA$^2$, in panel c,d is $200 \times 200$ \AA$^2$. The square in panel a indicates part of the area shown in b. Squares and the arrow in b indicate the areas shown in c and d.}
\label{fig:cu111_matsumoto}
\end{figure}

At room temperature the oxidation of Cu(111) also proceeds through the formation of epitaxial oxide islands with a $5\times6$ lattice misfit~\cite{zhou_apl_2009, leon, matsumoto_2001_ss} which coalesce with continuous oxygen exposure.
Three processes of oxide formation are observed in this regime~\cite{leon, matsumoto_2001_ss} (Fig.~\ref{fig:cu111_matsumoto}): growth from step edges, in-terrace growth ($\sim 1.2$ \AA\ below the terrace surface) from vacancy islands and growth of on-terrace oxide.
The formation of randomly placed islands at low temperature agrees well with the existence of a disordered underlying structure observed for low-coverage oxygen adsorption and analysed in Sec.~\ref{ocu111}.
At higher temperatures~\cite{zhou_apl_2009, matsumoto_2001_ss} fast formation of the oxide and fast lateral growth leads to the formation of a two-dimensional oxide structure.
This uniform growth is templated by the distorted-hexagonal surface reconstruction shown in Sec.~\ref{ocu111}~\cite{matsumoto_2001_ss}.

\begin{figure}
\centering
\includegraphics[width=100mm]{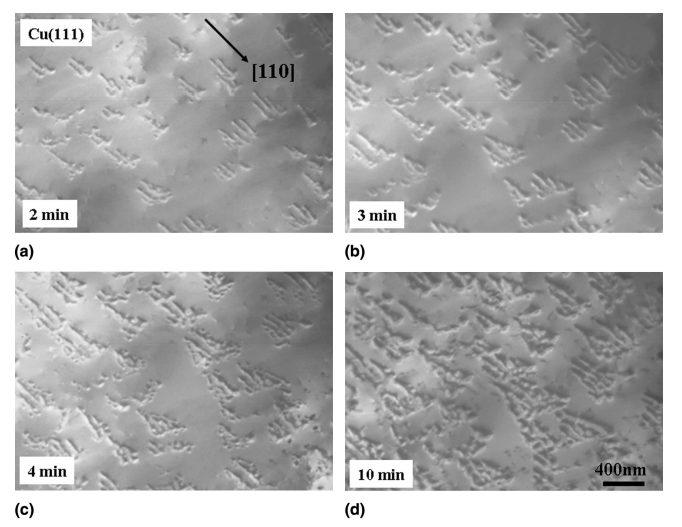}
\caption{Percolating oxide growth on Cu(111) observed using \emph{in situ} TEM imaging at $450$ $^{\circ}$C. 
The oxidation partial pressure is $P=5 \times 10^{-4}$ Tor. 
The layer is formed by preferential nucleation of islands along the [110] direction,
suggesting that between $350-450$ $^{\circ}$C the substrate transitions from a disordered
to an ordered structure. Taken from Ref.~\cite{zhou_jmr_2005}.}
\label{fig:cu111_perc}
\end{figure}

Zhou \emph{et al.} observed an interesting phenomenon at intermediate temperatures: the islands nucleate close to existing islands, anisotropically elongating along the [110] direction in a percolating manner, as shown in Fig.~\ref{fig:cu111_perc}.

This `discontinuous-branched' shape has been investigated using kinetic Monte Carlo techniques~\cite{zhou_apl_2008}, and it appears to be related to restricted diffusion of oxygen on the surface, which might then be related to the surface being reconstructed to the `29' or `44' structures, which are fairly corrugated (they extend up to $3.1$ \AA\ over the clean Cu(111) surface).

{\footnotesize
\begin{longtable}{|p{1.8cm}|p{1.5cm}|p{4cm}|p{2.5cm}|p{1.5cm}|p{3.5cm}|}
\hline
{\bf Reference} & {\bf Sample thickness} & {\bf Surface Preparation} & {\bf Exp. Conditions} & {\bf Technique} &{\bf Result} \\ \hline
Lawless~\cite{lawless_gwathmey} (1956) & -- & Electropolishing & $P=0.8-760$ Torr, $T=170-450$ $^{\circ}$C & XRD & Epitaxial Cu$_2$O oxide, disordered CuO. \\ \hline
Goulden~\cite{goulden} (1976) & $250-400$ $\mu$m & Polishing in orthophosphoric acid at 1.4 V & $8 \times 10^{-4}$ Torr, $T=250-400$ $^{\circ}$C & TEM & Epitaxial oxide islands. \\ \hline
Ho~\cite{ho_1978} (1978) & -- & Annealing in vacuum at $300$ $^{\circ}$C & $10^{-4}-10^{-7}$ Torr, $T=350^{\circ}$C & TEM & Epitaxial lamellar growth. \\ \hline
Dubois~\cite{dubois} (1982) & -- & Annealing in H$_2$ at $630$ $^{\circ}$C & $1 \times 10^{-3}$ Torr, $T=525$ $^{\circ}$C & EM & Epitaxial oxide islands on defect sites. \\ \hline
Milne~\cite{milne_howie} (1984) & $0.05$ $\mu$m  & Polishing, annealing & $P=10^{-5}-5 \times 10^{-4}$ Torr, $T=300$ $^{\circ}$C & RHEED, TEM & Formation of epitaxial oxide islands. \\ \hline
Rauh~\cite{rauh} (1993) & $50$ nm & Deposition by dc heating. & O$_2$, $P=7.5 \times 10^{-4}-9\times 10^{-2}$ Tor, $T=105$ $^{\circ}$C, $400$ min & Ellipsom. & Formation of a Cu$_2$O film \\ \hline
Matsumoto~\cite{matsumoto_2001_ss} (2001) & -- & Ar$^+$ sputtering and vacuum annealing at $500$ $^{\circ}$C & $P=10^{-7}-10^{-5}$ Torr, RT & STM, LEED & Growth of oxide from step edges. \\ \hline
Zhou~\cite{zhou_jap_2005} (2005) & $70-80$ nm & Annealing in CH$_4$O vapour at $350$ $^{\circ}$C/vacuum $800^{\circ}$C & $5 \times 10^{-4}$ Tor, $350$ $^{\circ}$C & TEM & Island nucleation rate faster on (111) than (110) or (100) \\ \hline
Zhou~\cite{zhou_jmr_2005} (2005) & $70-80$ nm  & Annealing in vacuum at $550$ $^{\circ}$C & $5 \times 10^{-5}$ Tor, $350-900$ $^{\circ}$C & TEM & Temperature-dependent shape and oxidation rate of the oxide islands. \\ \hline
Zhou~\cite{zhou_prl_2005} (2005) & $70-80$ nm  & Annealing in CH$_4$O vapour at $350$ $^{\circ}$C/vacuum $800$ $^{\circ}$C & $5 \times 10^{-4}$ Tor, $350$ $^{\circ}$C & TEM & Island nucleation rate faster on (111) than (110) or (100) \\ \hline
Zhou~\cite{zhou_apl_2008} (2008) & -- & Annealing in vacuum at $800$ $^{\circ}$C & $3 \times 10^{-4}$ Tor, $900$ $^{\circ}$C & TEM & Growth of oxide hollow pyramidal islands. \\ \hline
Zhou~\cite{zhou_apl_2009} (2009) & $60$ nm & Annealing in CH$_4$O vapour at $350$ $^{\circ}$C/vacuum $800$ $^{\circ}$C & $5 \times 10^{-4}$ Tor, $350$ $^{\circ}$C & TEM & Epitaxial Cu$_2$O $\sim2.5$ nm-high islands. \\ \hline
Leon~\cite{leon} (2012) & -- & Ar$^+$ sputtering and $550$ $^{\circ}$C heating & $P=10^{-7}$ Torr, RT, pulse injection of air & AES, STM & Oxide growth from step edges. \\ \hline
\caption{Experimental results of copper oxide formation in ultra-high vacuum on Cu(111). Experimental conditions and techniques have been listed, together with a summary of the main result of each study.}
\label{tab:uhv_111} 
\end{longtable}}

\subsection{Nano-island formation kinetics}
\label{controlled_4}

A number of models have been proposed to explain the kinetics governing the initial stages of oxide growth, from island nucleation to their coalescence.

\begin{figure}
\centering
\includegraphics[width=120mm]{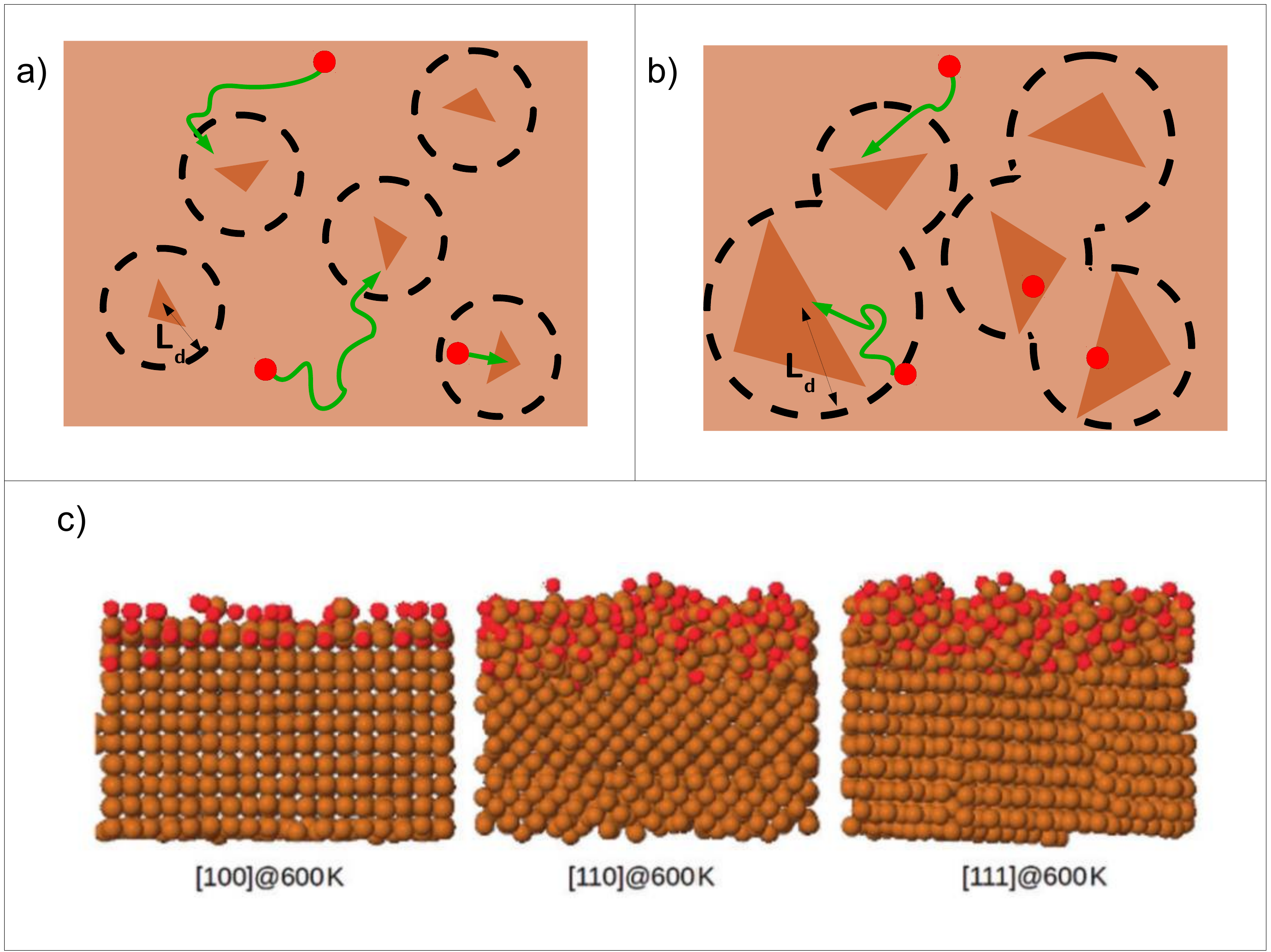}
\caption{a) Schematic representation of the mechanism of island nucleation and growth according to the model in Eq.~\ref{eq:yang}. The dashed circles represent the `oxygen capture zones' of radius L$_d$ around the oxide islands. Oxygen atoms (in red) diffuse on the surface until they enter a capture zone, when they are incorporated in an oxide island (represented as a brown triangle). b) Schematic representation of the mechanisms of island growth to coalescence. As the islands increase in size their areas of oxygen capture increase (black dashed line). Oxygen surface diffusion (green arrows) is still important, however direct impingement of oxygen atom in the islands is also more likely to occur. c) Oxidation of Cu(100), Cu(110) and Cu(111) copper surfaces using the ReaxFF force field at $T=300$ K and $T=600$ K.  Cu(111) is found to oxidise more easily than (110) and (100) and form a thicker oxide film. The amorphous nature of the formed oxide is clearly visible. Taken from Ref.~\cite{Byoungseon}.}
\label{fig:reaxff}
\end{figure}

The mechanism of nucleation and initial growth of oxide islands on a clean Cu surface has been proposed in terms of `capture zones', a well-established idea in non-homogeneous film formation theory~\cite{mulheran, evans}.
In this interpretation, schematically shown in Fig.~\ref{fig:reaxff}a, oxygen atoms landing in a capture zone, \emph{i.e.} an area with radius $L_d$ surrounding the island, will likely aggregate with the island and contribute to island growth rather than nucleating new islands.
Large `capture zones' (\emph{i.e.} large values of $L_d$) are associate with longer path lengths of surface oxygen diffusion.
In this framework, the saturation number of islands that can be nucleated, $N_{sat}$, is expressed as~\cite{yang_SM_1998}:
\begin{equation}
\label{eq:yang}
N_{sat}=\frac{1}{L^2_d}(1-e^{-kL^2_d t}),
\end{equation}
where $k$ is the initial island nucleation rate and $t$ is time.
This model was found to fit well experimental data for oxidation of Cu(100) and Cu(110)~\cite{yang_SM_1998, yang_mm_2001}, as shown in Table~\ref{table:model}.
\begin{table}[h]
\centering
\begin{tabular}{|c|c|c|c|}
\hline
      & $N_{sat}$ [$\mu \mathrm{m}^{-2}$] & $k$ [$\mu \mathrm{m}^{-2}\mathrm{min}^{-1}$] & $L_d$ \\ \hline 
(100) & 0.83 & 0.17 & 1.09 \\ \hline
(110) & 4.34 & 1.74 & 0.33 \\ \hline
\end{tabular}
\label{table:model}
\end{table}
The smaller number of islands nucleating on Cu(100) is related to a larger oxygen capture area, \emph{i.e.} longer path lengths for surface oxygen diffusion.
This is consistent with the structure of the missing-row reconstructed Cu(100) surface, which is fairly smooth (it has a corrugation of 0.35 \AA~\cite{jensen_prb_1990}), see Sec.~\ref{ocu100}, thus favouring surface diffusion.
Similarly, the short diffusion path length for Cu(110) is consistent with the added-row reconstruction which protrudes $\sim1.5$ \AA\ over the Cu surface~\cite{tanaka,besenbacher} (see Sec. \ref{ocu110}).
Nucleation is seen to be faster on Cu(110) than on Cu(100), see Table~\ref{table:model}, and even faster on Cu(111) at $350^{\circ}$C~\cite{zhou_jmr_2005}.
The O-induced reconstruction of the Cu(111) surface has been reported to have a disordered surface overlayer at room temperature, with O and Cu atoms at different heights~\cite{jensen1,jensen2} (see Sec~\ref{ocu111}).
This could explain a shorter path length for oxygen diffusion on Cu(111) leading to a fast nucleation of a large number of oxide islands.

After island nucleation has reached saturation point, they start growing until coalescence.  
Island growth models based on oxygen impingement and surface diffusion (schematically represented in Fig.~\ref{fig:reaxff}b) have been developed~\cite{orr, holloway_hudson, yang_apl_1997}: oxygen surface diffusion initially dominates the oxide growth, and later oxygen direct impingement becomes significant when the oxide islands grow larger in size.
In particular the model proposed by Yang \emph{et al.}~\cite{yang_apl_1997} was found to fit well experimental data on Cu(100), with island growth proportional to  $t^{1.3}$.

Another well established theory for the formation of thin films is the Avrami (or JMAK) nucleation model~\cite{kolmogorov,avrami,johnson}.
It presumes isotropic and homogeneous nucleation of the islands and depends exponentially on time:
\begin{equation}
X(t)=1-e^{-kt^n},
\end{equation}
where $X$ is the oxide thickness.
Cu(100) fits this model~\cite{yang_mm_2001} with $k=1.9 \times 10^{-4}$ and $n=2$.
The value of $k$ is much smaller than expected (typically it is $k=\pi/3$), possibly because of the non-constant island nucleation in cuprous oxide (but rather following the relation in Eq.~\ref{eq:yang}) and the non-linear island growth rate ($A(t) \propto t^{1.3}$).
Yang {\it et al.} modified the JMAK model in order to take into account these two factors and found an excellent fit with Cu(100) experimental data~\cite{yang_apl_2002}.

Although the nucleation of islands is faster on Cu(110), the long term ($> 60$ minutes) oxidation of the (100) surface of copper is much faster than on the (110) and (111)~\cite{lawless_gwathmey, young, rhodin} surfaces.
Fast initial nucleation and growth of islands on Cu(110) and Cu(111) leads quickly to a thinner coalesced film, which then continues growing through oxygen diffusion through the oxide layer, a much slower process than surface diffusion. 
On the contrary, the slow nucleation of islands on Cu(100) leads to slower coalescence to a thicker oxide film with respect to the Cu(110) and Cu(111) and thus quicker oxide growth by means of surface diffusion.

The case of oxidation on Cu(111) at higher temperatures (over $550$ $^{\circ}$C) is the only one where Cu$_2$O two-dimensional thin film growth is observed~\cite{zhou_jmr_2005, leon}.
The O-induced surface reconstruction of Cu(111) at high temperatures, as seen on Sec.~\ref{ocu111}, has a hexagonal morphology similar to the Cu$_2$O(111) plane~\cite{jensen1,jensen2} which can act as a `template' structure for the layered growth of the oxide.

A few computational studies have tried to obtain further insight into the kinetics of oxide growth, with limited success so far.
The correct relative rates on the three low-index Cu surfaces have been obtained using molecular dynamics with ReaxFF, a bond order potential~\cite{Byoungseon}.
However the calculation led to the formation of a uniform amorphous thin film (as shown in Fig.~\ref{fig:reaxff}c), thus not reproducing crystalline island formation as seen in experiments, and the role of the surface reconstruction was not taken into account.
As already mentioned, Devine \emph{et al.}~\cite{devine_prb_2011}, looked to reproduce oxidation on an O-reconstructed Cu(100) surface, using molecular dynamics and a bond order potential.
However, the process was too slow and the simulation too short for it to be modelled.

\subsection{Long-term copper oxidation}
\label{controlled_1}

\begin{figure}
\centering
\includegraphics[width=120mm]{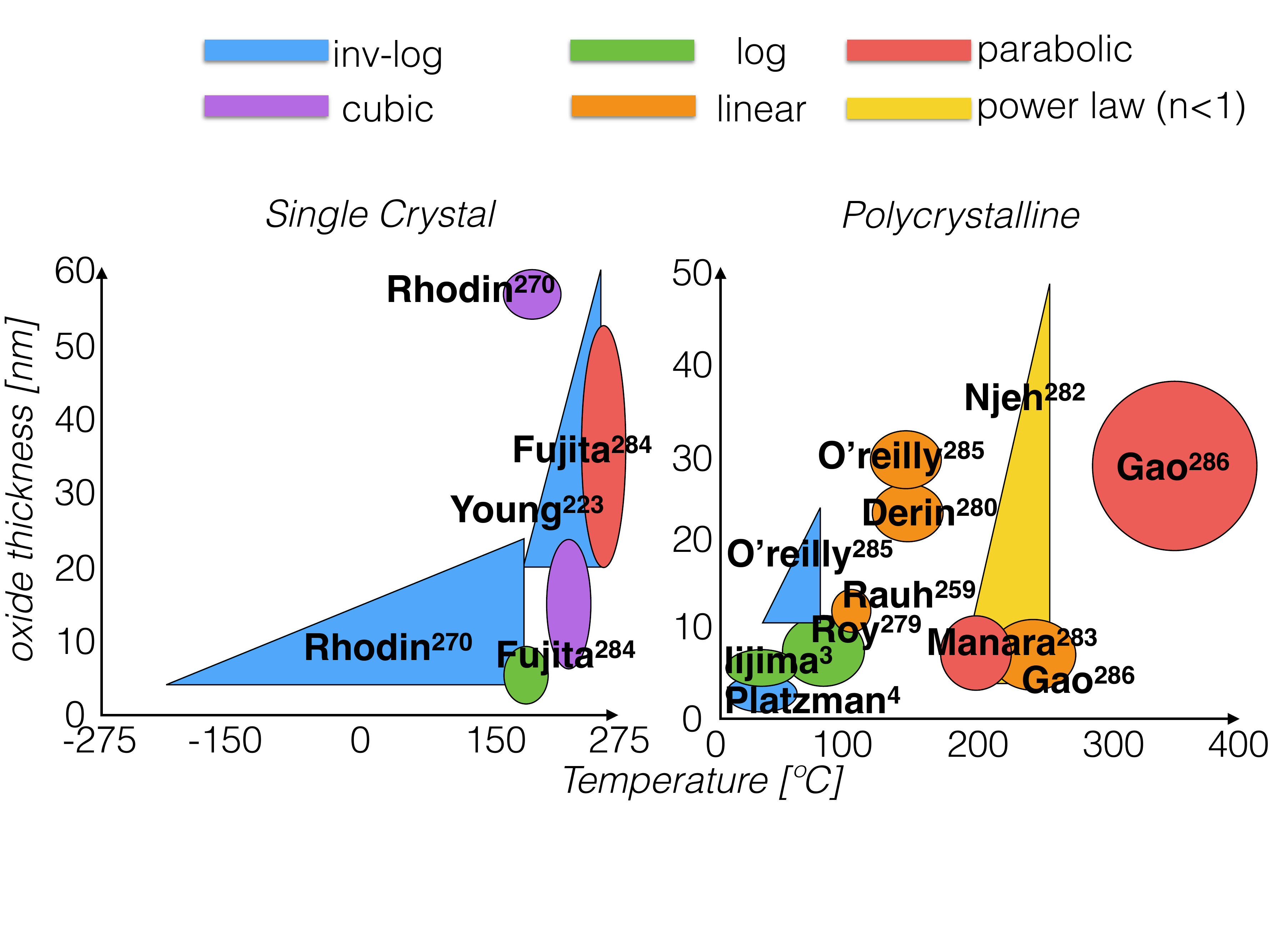}
\caption{Overview of growth models for copper oxide present in the literature. The graphs show the oxide film thickness vs oxidation temperature for a number of experimental studies. The data have been divided into two categories, according the composition of the initial copper sample, single crystal or polycrystalline. They have also been colour coded according to the oxidation kinetic model which has been attributed to them. It can be seen that on the basis of these three parameters only (temperature, film thickness and initial sample structure) a single kinetic model cannot be identified.}
\label{fig:papers}
\end{figure}

After nucleation, growth and coalescence of copper oxide islands, oxidation proceeds through the diffusion of oxygen atoms through the oxide and the character of the oxidation process changes.

The kinetics and mechanism of copper oxidation, after the exposure of the copper surface to oxygen flow for an extended period (up to 32 hours) have been extensively studied~\cite{lawless,leidheiser1971corrosion, ronnquist_review}, however no consensus on a single rate law describing the oxidation dynamics has been established.

Clean copper surfaces have been found to oxidize at different rates, with the Cu(100) face reported to have the fastest rate of oxidation, at odds with the oxidation rates
measured for the formation and growth of oxide nano-islands.
Indeed, Young~\cite{young} \emph{et al.} and Gwathmey~\cite{gwathmey} \emph{et al.} found that the order of oxidation rate for the low-index surfaces is (100), (111), (110) with (100) the fastest oxidising facet, for a wide range of temperatures. 
Rhodin~\cite{rhodin_2} found instead the order to be (100), (110), (111) with (100).
This difference in ordering could be due to several factors, for example the use of very different experimental analysis techniques.
Gwathmey~\cite{gwathmey} \emph{et al.} used diffuse white light to analyse the oxidized surfaces, observing different interference colours corresponding to different oxide thicknesses, Young~\cite{young} \emph{et al.} used ellipsometry which relies on a `guess' of a surface model and Rhodin~\cite{rhodin_2} calculated the film thickness from weight changes in the sample.
All three methods present some drawbacks (diffuse white light does not provide a value for the thickness and relies on visual checking of the surface colour, ellipsometry relies on a guess model and Rhodin's method depends on an accurate estimation of the effective area of the resulting oxide) and it is difficult to tell which of the three is more reliable.
The difference observed could also be the result of environmental factors such as the presence of impurities or different sample preparations, which have been shown to lead to different resulting metal oxides~\cite{bouzidi_ss_2005, lim} which makes oxidizing metals very challenging to study.

Very large discrepancies are observed in models for oxidation rates, \emph{i.e.} growth of thickness of the oxide as a function of time, and they are summarised in Fig~\ref{fig:papers}.
A number of theories have been proposed~\cite{landsberg_jcp_1955, grimley_prsl_1956, uhlig_am_1956, evans_tes_1947}, mostly based on the Cabrera and Mott~\cite{cabrera} theory and they postulate that, under the assumption of uniform epitaxial growth, the thickness of a metal oxide increases following an inverse logarithmic rate law for very thin films (up to $7.3$ nm for Cu) and a cubic law for thicker films (up to $1.5$ $\mu$m).
A number of works~\cite{rhodin_2, hu_2000, young, roy_om_1981} find qualitative agreement with the Cabrera-Mott theory, however linear oxide growth has been observed in other studies~\cite{rauh, derin_apa_2002, vanwijk}, as well as power ($n<1$)~\cite{njeh_sia_2002} or parabolic law~\cite{manara_sia_1992}.

Many factors can affect the experimental measurements of oxide kinetics, such as the environment and the type of initial copper sample.
Indeed, it was found that polycrystalline film oxidation kinetics is almost twice as slow as single crystal oxidation kinetics~\cite{vanwijk}.
The oxidation temperature is another factor~\cite{fujita_ass_2013, oreilly, gao_ml_2001}.
O'Reilly~\cite{oreilly} \emph{et al.} found for a polycrystalline sample in dry synthetic air (O$_2$/Ar mixture) that at temperatures between $250$ and $500$ $^{\circ}$C the oxidation followed a cubic law, at $100$ $^{\circ}$C an inverse log rate and at $150$ $^{\circ}$C a linear growth rate.
Gao~\cite{gao_ml_2001} \emph{et al.} found a linear oxidation rate between $200-250$ $^{\circ}$C with a fine-grained Cu$_2$O being the resultant oxide, and parabolic above that, with CuO the final oxidation product.

Although a lot of work has been performed in this field, it is clear from Fig.~\ref{fig:papers} that there is no consistency in the results obtained.
Systematic work looking at the influence of crystal structure (single crystal, polycrystalline, different grain size), of the experimental set-up (humidity, temperature, oxygen pressure) and composition of the oxide product need to be performed in order to clarify where the contributions to these different oxidation rates originate from.

\subsection{Native oxidation of copper in ambient conditions}
\label{native}

Structural details of the room temperature oxidation of copper under ambient conditions, including the possible influence of humidity and crystal structure, are not very clear. 

In order to study copper in ambient conditions (room-temperature, 1 atm pressure) precise measurement techniques are required, since surface oxide films are generally only a few nm thick. 
Experiments studying oxidation in these conditions have been performed for an enormous range of exposure times, between 30 minutes to 9 months, in air at ambient conditions (referred to as ambient air hereafter), obtaining results which are not always consistent (see Table~\ref{table:native}). 
As we will see, the outcome of the studies differs in the thickness of the final oxide, in the presence or not of a CuO overlayer on top of the  Cu$_2$O oxide layer and in the 
order of formation of the oxides. 
The reason for this non-uniformity is probably to be found in the method of preparation of the film, on the type of film used (thin/bulk, polycrystalline/single crystal), the 
ambient humidity (which is rarely reported) and the surface roughness.

Native oxidation has been extensively studied in copper thin films (of the order of a few hundred nms), because of the important role it plays in the passivation of nano-sized electronic copper parts, using XAS and XPS to identify the composition of the top layer and ellipsometry to measure the film thickness.
Platzman \emph{et al.}~\cite{platzman_2008} proposed a three-stage oxidation mechanism involving: (a) the formation of a Cu$_2$O layer, most likely due to Cu metal ionic transport toward the oxide-oxygen interface; (b) the formation of a Cu(OH)$_2$ metastable overlayer, due to the interactions of Cu ions with hydroxyl groups present at the surface; and (c) the transformation of the Cu(OH)$_2$ metastable phase to a more stable CuO layer.
Indeed, the formation of a $\sim 2.0-5.0$ Cu$_2$O layer first, followed by a $\sim{0.9-1.3}$ CuO  overlayer has been reported~\cite{platzman_2008, iijima, keil2010, keil2007}.
However, Lim \emph{et al.}~\cite{lim} showed that the texture and microstructure of a thin copper film have a direct influence on the oxidation products. 
When oxidising a sample with a columnar structure and small grains the Cu$_2$O/CuO bylayer was obtained, when oxidising a uniform sample with no obvious columnar structure and grain boundaries the cuprous oxide layer only was observed and oxidation was slower. 

The oxidation of bulk copper studies was studied by means of XPS, XRD and ellipsometry, and the results obtained are mixed.
A number of studies~\cite{iijima, chawla, chu} found only the formation of $\sim1.6$ nm thick Cu$_2$O even after long exposure times. 
However, other studies of passivated Cu after exposure at room-temperature air~\cite{barr1, suzuki} observed the formation of a few ML of CuO, which was found to start growing only after the  Cu$_2$O growth process has finished. 
The different results obtained by these experiments can be due to the texture and microstructure of the 
copper film~\cite{lim}, by the surface roughness or by defects present at the surface.
Indeed, the surface roughness of metal oxide surfaces has been shown to have a direct influence on their 
wetting properties towards water, which in turn could have a direct influence on the formation of the native 
oxide~\cite{rico}. 
In particular, very smooth copper surfaces are hydrophobic while rough surfaces ($\sim 5$ nm-high roughness)
are hydrophilic~\cite{platzman_langmuir}. 
Furthermore the orientation of the crystallites at the surface influences the wetting properties and the 
distribution and coverage of water on the surface, especially in the case of copper~\cite{yamamoto}. 
Moreover, polycrystalline structures with nanosized grains have higher surface energy at the grain boundaries 
than structures that are made of micrometric grains or crystalline lattice, which in turns affect the wettability of the surface.


\newpage
\begin{landscape}
\small
\begin{longtable}{|p{1.7cm}|p{2.5cm}|p{3cm}|p{3.3cm}|p{2cm}|p{2.7cm}|p{5cm}|}
\hline
{\bf Author} & {\bf Cu thickness} & {\bf Environment} & {\bf Surface treatment} & {\bf Exposure time} & {\bf Technique} & {\bf Resulting oxide structure} \\ \hline
Barr~\cite{barr1} (1978) & Commercial  film & RT air, $35\%$ humidity & Sputter etching & $30$ min/days & ESCA spectrometry & $2.0$nm Cu$_2$O/a few ML  CuO-$2.0$ nm Cu$_2$O \\ \hline
Chawla~\cite{chawla} (1992) & Commercial film & Air & Electropolishing & 24h in a desiccator & XPS & $1.6$ nm  Cu$_2$O layer. \\ \hline
Chu~\cite{chu} (1999) & Commercial film & RT air & Electropolishing & 23h & XRD & Cu$_2$O only (both in air and in solution) \\ \hline
Iijima~\cite{iijima} (2006) & Cu bulk 0.5mm/ Cu film & RT air & Electrochemical polishing, annealing in dry H at 600°. & 30 min- 1300h & XPS/ ellipsometry & Cu$_2$O only on Cu bulk, outer CuO film on the Cu thin film \\ \hline
Keil~\cite{keil2007,keil2010} (2010) & 90 nm & RT air & DC sputtering on glass. & 48h/9months & XAS & Oxide bilayer: 2.0 nm CuOover  3.5 nm Cu$_2$O \\ \hline
Lim~\cite{lim} (2008) & 100 nm & RT air, $10\%$ humidity & ion beam deposition, argon ion sputtering & 30 min/1280h & XPS, SEM, HRTEM & Cu deposited with $0$ V bias voltage at the substrate shows both  Cu$_2$O and CuO, with $-50$ V only  Cu$_2$O.\\ \hline
O'Reilly~\cite{oreilly} (1995) & bulk (2mm), film:80-500 nm & T=50-150, dry synthetic air. & Bulk: ground surfaces. Film: electroless deposition, sputtering. & 300 min & TGA, XRD & $T< 100$ $^{\circ}$C, only Cu$_2$O, $T> 100$ $^{\circ}$C, Cu$_2$O and CuO.\\ \hline

Platzman~\cite{platzman_2008} (2008) & $\sim400$nm & RT air, $60\%$ humidity. & Thermal filament evaporation deposition on Si wafers. & 1h-112 days & XPS, TEM, SEM & After 1h only Cu$_2$O observed, CuO starts occurring after 24h. \\ \hline

Suzuki~\cite{suzuki} (1997) & 0.5mm & RT air & Electrolytical polishing & 10min-12d & XPS & $2^{+}$ and $1^{+}$ oxidation states present.\\ \hline

\caption{Summary of experimental studies on the native oxidation of copper. Details of the sample type, preparation and environmental conditions have been specified, together with the experimental techniques used and the observed oxide structure.}
\label{table:native}
\end{longtable}
\end{landscape}

\newpage

\subsection{Nucleation sites}
\label{controlled_5}

A number of studies have tried to establish the role of surface defects on the nucleation of the oxide, in order to understand whether island nucleation is a heterogeneous process, triggered by specific surface features, or a homogeneous process.
While some metals and semimetals can grow oxide layers homogeneously without the aid of impurities or surface defects (such as Ru(0001)~\cite{bottcher_jpc_1999}, Ag(110)~\cite{bao_cl_1995} 
or Si(111)~\cite{ross_prl_1992, dujardin_prl_1996}) it has been shown that other metals, like Pb~\cite{thurmer_science_2002} or Ge(111)~\cite{dujardin_prl_1999, mayne_ss_2003}, cannot oxidise without the presence of surface features (impurities or defects) which trigger the dissociation of O$_2$ molecules.

For copper it has been shown~\cite{dubois, zhou_jap_2005, zhou_cc_2013, matsumoto_2001_ss} that defect sites play a role in oxide island nucleation.
Grain boundaries~\cite{zhou_jap_2005}, vacancy islands~\cite{matsumoto_2001_ss} and the edges of pits~\cite{yang_jes_1999} are found to be nucleating sites for island formation, however no preferential nucleation sites have been found at dislocations, stacking faults or impurities~\cite{yang_jes_1999, heinemann, zhou_jap_2005}.
The importance of step edges was initially inferred by Milne and Howie~\cite{milne_howie}, and it was then demonstrated on Cu(111) by means of STM and TEM~\cite{leon, matsumoto_2001_ss}.
However, TEM work by Yang~\emph{et al.}~\cite{yang_jes_1999, zhou_jap_2005} showed that this is not the case on Cu(100) and Cu(110) films.
It is indeed possible, considering the different nature of oxide formation on the three low-index surfaces, that different defects play a more or less important role in different oxide nucleation conditions.

More work is needed to clarify further the correlation between defect sites and oxide formation and remove doubts on whether small, non-structured dislocations and impurities have a role in the nucleation of copper oxide islands, and, from a theory point of view, why some type of defects seem more efficient at nucleating oxide islands than others.

\subsection{CuO formation}
\label{controlled_cuo}

CuO is expected to form after exposure of a copper surface to oxygen at high temperatures and pressure~\cite{honjo,eastman}.
Much work has gone into understanding high temperature oxidation of copper ($>350^{\circ}$) and it has been recently reviewed~\cite{zhu_mmta_2006, belousov_rcr_2013}.
Most work shows the growth of CuO on top of Cu$_2$O, following a parabolic rate law for the thickness as a function of time.
However, no studies of the atomistic details of the formation of CuO at high temperatures have been performed to date.

In controlled conditions, evidence of a CuO overlayer was found by exposing a Cu sample to a controlled flow of O$_2$ at high pressure and the dependence of oxidation on oxygen pressure was analysed.
Boggio~\cite{boggio} investigated the pressure dependence ($P=0.03-7.5$ Torr) on the oxidation of Cu(111), and in particular the film thickness, using ellipsometry. 
The film growth (thought to be Cu$_2$O), after $90$ minutes exposure at $21^{\circ}$, was related to the Cabrera-Mott expression of growth. 
However, a dramatic decrease in the oxidation rate with increased oxygen pressure was observed and related to the formation of a passivating CuO film at the oxide-oxygen interface.
Pierson~\cite{pierson} \emph{et al.} looked at the reactivity of a number of noble metals when subject to a flow of gases. 
In the case of Cu in an O$_2$ flow,  the formation of oxidised structures was found to depend on the flow rate of O$_2$. 
The formation of Cu$_2$O (at oxygen partial pressure p(O$_2$)=$7.5 \times 10^{-5}$ Torr), CuO (p(O$_2$)=$1.12 \times 10^{-3}$ Torr) and metastable Cu$_4$O$_3$ (p(O$_2$)=$1.5 \times 10^{-4}$ Torr), was observed by XRD analysis.

\section{Conclusion and discussion}
\label{conclusions}

The oxidation of copper and the physical properties of the resulting oxides are fundamental scientific problems which are still not completely understood today.
Since there has currently been a surge of interest in the use of copper oxide for catalysis, optoelectronics and gas sensing, the need for a detailed understanding of the surface structures of these oxides has become even more pressing.
In addition, uncontrolled copper oxidation is still an issue in \emph{e.g.} electronic applications, and understanding the oxide growth process is the first step towards mitigating it.

In this review we have discussed the state of the knowledge regarding the structure and formation of copper oxides and we have seen that the structural, optic and vibrational characteristics of bulk copper oxides are well understood.
A good amount of computational work (mainly from DFT) has been performed on both Cu$_2$O and CuO, providing important information on the structures of the surfaces at different temperature and pressure conditions.
A hexagonal structure presenting Cu surface vacancies was found to be the most stable on Cu$_2$O and the stoichiometric (111) surface was the most stable for CuO.
Few experimental studies are available to either confirm or disprove these suggestions from theory.
Considering the potential technological applications of these oxides, this is an area which should be looked at more with experimental techniques such as STM, XPS or LEED.

Many atomistic aspects of the formation of the oxides have been established.
Indeed, the oxygen-induced surface reconstructions on copper surfaces with low Miller index are well known.
In addition, STM studies have revealed certain atomic level details of the initial stages of Cu surface oxidation.
Other experiments at low oxygen pressure show that the initial oxide growth happens through the formation of cuprous oxide islands which eventually coalesce.
The kinetics of the oxide formation depends on the temperature and on the copper surface and it is dominated by oxygen surface diffusion and direct impingement in the first instance, and after coalescence by oxygen diffusion into the bulk.

There is however scope for further work as many fundamental aspects of copper oxidation are still unclear.
In fact, there is no consensus on how the transition from oxygen-reconstructed copper surfaces to the onset of 
oxidation occurs, whether this is after the surface has been reconstructed or if the islands start forming on the clean surface.
At least for the case of Cu(100) and Cu(110), there is good evidence that the onset of oxidation will occur after the formation of an O-reconstructed overlayer which affects the growth kinetics of the islands themselves.
The precise understanding of the formation of these copper islands is not only important to understand the onset of oxidation, but also to exploit the oxide islands as nano-templates for technological applications.
In order to do that, the precise mechanism of nucleation needs to be understood.

It is also unclear where the different behaviour between metals such as Ag or Al, which grow uniform oxide layers, and copper stems from.
The O-induced surface reconstructions of different fcc metals have been extensively studied~\cite{besenbacher} and the 
differences found in their electronic structures~\cite{courths_ss_1997} or surface configurations~\cite{schimizu} could provide 
hints to their different behaviour upon higher exposure to oxygen.
However, dedicated theoretical studies to this aim have not been performed so far.

The kinetics of long-term oxide growth, both in controlled conditions and in ambient air is an issue that remains poorly understood.
Copper oxide formation does not simply follow the Cabrera-Mott law, especially at the initial stages when oxide growth is not uniform across the surface.
Many studies have been performed in order to establish the oxide formation kinetics at low temperatures,
however they are difficult to compare to one another, because of the different conditions in use.
Indeed temperature, humidity, oxygen partial pressure, structure of the initial copper film/coupon, the presence of defects and impurities
seem to affect the growth of the oxide as well as the final oxide product.
Moreover, the structure of copper oxide when grown in ambient air, which is very important to the practical 
uses of copper, is still not clearly understood, with the formation of a CuO overlayer on top of a Cu$_2$O layer still being debated.
Further systematic experimental studies, looking at disentagling the environmental factors influencing oxide growth are needed.
Computationally, additional studies of copper and copper oxide surfaces and their interaction with the environment (\emph{e.g.} water, O$_2$, N$_2$) would provide valuable information in support of the experimental work.
Work in this direction has recently been reported, \emph{e.g.} water and hydroxide adsorption on Cu and Cu oxide~\cite{forster_cs_2012, kronawitter_jacs_2014, deng_lang_2008, jiang_ec_2010}.

Part of the challenges that have been faced by scientists in studying these systems are due to the limitations of theoretical and experimental technology available to them.
It is indeed clear that massive steps forward have been made since the first calorimetry experiments on Cu$_2$O crystals, and nanometre scale resolution has been obtained with STM and TEM and structural atomistic structures can be explored with XPS and LEED.
Moreover, development of experimental techniques, such as X-ray lasers with extremely high spatial and temporal resolution~\cite{nilsson_science} or near-ambient pressure photoelectron spectroscopy~\cite{nap-pes}  might enable mechanisms of initial stages of oxidation to be explored.

Computational techniques also present challenges from the point of view of accuracy of calculated structures and physical properties, as well as time length of molecular dynamics simulations.
Moreover developments in theory, especially \emph{ab initio} molecular dynamics, accelerated sampling techniques and more sophisticated non \emph{ab initio} approaches mean that similar studies on the first steps of oxidation are possible.
\emph{Ab initio} molecular dynamics, along with a free energy sampling approach has been used to examine the initial stages of NaCl dissolution in liquid water~\cite{nacl_diss}: it is not inconceivable that similar techniques could be applied to copper oxidation in ambient and even aqueous conditions.

Finally, another important challenge, is bridging the gap between highly controlled studies performed in ultra-high vacuum and work aiming to understand the formation of the oxide in industrially relevant conditions.
At the moment, these two aspects of the oxidation problem are addressed using different techniques at different resolutions.
In order to fully understand the oxidation of copper it is important to be able to relate the results obtained in these different conditions to one another and build a unified picture of the problem.


\section*{Acknowledgements}

A.M.'s work is partly supported by the European Research Council under the European Union's Seventh Framework Programme (FP/2007-2013)/ERC Grant Agreement No. 616121 (HeteroIce project) and the Royal Society through a Wolfson Research Merit Award.

\footnotesize{
\bibliography{oxid_copper_11} 
\bibliographystyle{rsc}
}

\end{document}